\documentclass[showpacs,floatfix,citeautoscript,twocolumn,prb,superscriptaddress]{revtex4-1}

\usepackage{graphicx,xcolor}
\usepackage{amsmath}
\usepackage{comment}
\usepackage{textcomp}
\usepackage{inputenc}

\newcommand*{\cumnas}{Cu$_{0.82}$Mn$_{1.18}$As}
\newcommand*{\cuexcess}{Cu$_{1.18}$Mn$_{0.82}$As}
\newcommand*{\mnexcess}{Cu$_{0.64}$Mn$_{1.36}$As}
\newcommand*{\asexcess}{CuMn$_{0.964}$As$_{1.036}$}

\newcommand{\mhk}[1]{\textbf{\textcolor{red}{#1}}}

\begin{document}

\title{High-resolution diffraction reveals magnetoelastic coupling and coherent phase separation in tetragonal CuMnAs}

\author{Manohar H.\ Karigerasi}
\author{Kisung Kang}
\author{Jeffrey Huang}
\affiliation{Department of Materials Science and Engineering and Materials Research Laboratory, University of Illinois at Urbana-Champaign, Urbana, IL 61801, USA}
\author{Vanessa K. Peterson}
\author{Kirrily C. Rule}
\author{Andrew J. Studer}
\affiliation{Australian Nuclear Science and Technology Organisation, New Illawarra Road, Lucas Heights, NSW 2234, Australia}
\author{Andr\'{e} Schleife}
\affiliation{Department of Materials Science and Engineering and Materials Research Laboratory, University of Illinois at Urbana-Champaign, Urbana, IL 61801, USA}
\affiliation{National Center for Supercomputing Applications, University of Illinois at Urbana-Champaign, Urbana, IL 61801, USA}
\author{Pinshane Y. Huang}
\author{Daniel P. Shoemaker}\email{dpshoema@illinois.edu}
\affiliation{Department of Materials Science and Engineering and Materials Research Laboratory, University of Illinois at Urbana-Champaign, Urbana, IL 61801, USA}


\begin{abstract}
Tetragonal CuMnAs was the first antiferromagnet where reorientation of the N\'eel vector was reported to occur by an inverse spin galvanic effect.
A complicating factor in the formation of phase-pure tetragonal CuMnAs is the formation of an orthorhombic phase with nearly the same stoichiometry. Pure-phase tetragonal CuMnAs has been reported to require an excess of Cu to maintain a single phase in traditional solid state synthesis reactions. 
Here we show that subtle differences in diffraction patterns signal pervasive inhomogeneity and phase separation, even in Cu-rich \cuexcess.
From calorimetry and magnetometry measurements, we identify two transitions corresponding to the N\'eel temperature (T$_N$) and an antiferromagnet to weak ferromagnet transition in \cuexcess\ and \asexcess. These transitions have clear crystallographic signatures, directly observable in the lattice parameters upon in-situ heating and cooling. The immiscibility and phase separation could arise from a spinoidal decomposition that occurs at high temperatures, and the presence of a ferromagnetic transition near room temperature warrants further investigation of its effect on the electrical switching behavior.

\end{abstract}

\maketitle 

\section{Introduction}

In 2016, Wadley \emph{et al}. \cite{Wadley2016} showed that it is possible to switch Mn moments in tetragonal CuMnAs between $[100]$ and $[010]$ using electrical currents. Metallic antiferromagnets, such as CuMnAs and Mn$_2$Au, are globally centrosymmetric but locally non-centrosymmetric and the individual sublattices are related to each other by an inversion center \cite{Zelezny2014,Wadley2016,Zelezny2017}. Since then, there has been a growing interest in studying the magnetic ordering in metallic antiferromagnets \cite{Wadley2013,Wadley2015,Hills2015,Saidl2017} and understanding how to read and manipulate their order parameter \cite{Wadley2016,Grzybowski2017,Wadley2018,Matalla-Wagner2019}. Unlike spin transfer torque based switching \cite{Gomonay2010}, the fieldlike torque from the inverse spin galvanic effect does not require an adjacent ferromagnet (FM) polarizer. This provides opportunity to synthesize bulk stress-free samples that do not require a substrate for the measurements. Large single crystals are also required for studying magnetic anisotropy using inelastic neutron scattering techniques\cite{Karigerasi2020}. However, all attempts to grow bulk crystals of CuMnAs so far have only resulted in $\mu$m-sized grains \cite{Volny2020}. This is in contrast to compounds such as Fe$_2$As, which has the same structure type as CuMnAs and can be grown in cm-sized single crystals \cite{Karigerasi2020}.

Complex structural and magnetic phase behavior is crucial to understand in the Cu-Mn-As system: subtle effects from antiferromagnetic domains can be easily overwhelmed by small ferromagnetic moments, and other magnetic orderings (Mn$_2$As in particular) do not possess the same symmetry as ideal CuMnAs. We show here that ferromagnetic moments are intrinsic in Mn-rich CuMnAs materials.
Bulk ternary compounds in the Cu-Mn-As system can be grown using traditional solid state synthesis routes \cite{Uhlirova2015,Uhlirova2019,Karigerasi2019}. However, when Cu, Mn, and As elemental powders are mixed in stoichiometric proportions, the orthorhombic polymorph of CuMnAs is stabilized \cite{MacA2012,emmanouilidou_magnetic_2017,Emmanouilidou2019}. 
Typically, substituting small amounts of Mn with Cu helps in stabilizing the tetragonal phase and the crossover from the orthorhombic to tetragonal phase for Cu$_{1+x}$Mn$_{1-x}$As lies somewhere between $x = 0.06 - 0.11$ \cite{Uhlirova2019}. Near-stoichiometric tetragonal CuMnAs can also be synthesized by substituting Mn with As \cite{Uhlirova2019} and electrical transport studies have been carried out on devices prepared from bulk samples \cite{Volny2020}. On the Mn-excess side of Cu$_{1+x}$Mn$_{1-x}$As, the thermodynamically stable phase changes from orthorhombic CuMnAs\cite{Uhlirova2015} to hexagonal \cumnas\cite{Karigerasi2019} to orthorhombic CuMn$_3$As$_2$\cite{Uhlirova2015} and eventually tetragonal Mn$_2$As \cite{Austin1962}. Cu is known to substitute up to $x = 0.1$ in Mn$_{2-x}$Cu$_x$As \cite{Uhlirova2019}. The crossover to tetragonal Mn$_2$As is intriguing since it has the same structure type as tetragonal CuMnAs. It may be possible, therefore, to also synthesize phase pure tetragonal CuMnAs with Mn excess.

The effect of increasing Cu substitution results in a decrease of the N\'eel temperature (T$_N$) \cite{Uhlirova2019}. Separately, an AFM-FM transition was observed by Uhlirova, et al.\ in near-stoichiometric samples around 315~K which was attributed to a possible MnAs impurity \cite{Uhlirova2019}. The first manuscript on the discovery of CuMnAs also reports a Curie temperature (T$_c$) at 300~K \cite{Nateprov2011}. 
Surprisingly, the AFM-FM transition has not been reported in any of the thin-film electrical switching papers \cite{Wadley2016,Meinert2018,Matalla-Wagner2019}. 
Confirmation of the absence of a FM component in thin film transport studies is not routine, so the possibility of coupling to a polarizable moment must be investigated, and we confirm this AFM-FM transition with diffraction and calorimetry measurements here. Despite the significance of bulk CuMnAs in understanding spin orbit torques in AFM, most electrical switching studies have only used epitaxially grown thin films. While magnetometry and calorimetry measurements have been carried out for the tetragonal phase on the Cu-excess side of Cu$_{1+x}$Mn$_{1-x}$As,\cite{Uhlirova2019} high resolution x-ray and neutron diffraction measurements are warranted for studying the phase stabilities and for understanding anomalies such as the low temperature ferromagnetic transition,\cite{Uhlirova2019,Nateprov2011} unipolar magnetic anisotropies, and low anisotropic magnetoresistance values \cite{Volny2020}.

In this article, we synthesize Cu-Mn-As samples at three different stoichiometries: Cu-rich \cuexcess, Mn-rich \mnexcess\ and a near-stoichiometric \asexcess. The near-stoichiometric tetragonal \asexcess\ composition has been reported and studied in previous papers (Uhlirova et al. (2019), Volny et al. (2020) etc.),\cite{Uhlirova2019,Volny2020}  so we synthesize and examine it here. Our Cu-excess \cuexcess\ composition lies close to the boundary between the tetragonal and orthorhombic structure (x = 1.11 in Uhlirova et al. (2019)).\cite{Uhlirova2019} Our Mn-excess composition \mnexcess\ is investigated here because it is the first compound along the Cu$_x$Mn$_{1-x}$As line that is Mn-rich and showed a pure tetragonal structure, without traces of orthorhombic CuMnAs.  Using scanning transmission electron microscopy (STEM) imaging, synchrotron x-ray diffraction (XRD), and neutron powder diffraction (NPD) measurements, we examine complex phase separation and sample inhomogeneity in the Cu-rich and Mn-rich samples.
Using calorimetry, superconducting quantum interference device (SQUID) magnetometry, synchrotron XRD and NPD measurements, we report strong magnetoelastic transitions at around 300~K and $T_N$ in the Cu-rich and the near-stoichiometric samples. 
The coherent stripe order of alternating domains with different Cu/Mn ratios implies that they could be altered by thermal cycling and likely contribute to anisotropic magnetoresistance measurements.

\section{Methods}

All three samples, Cu-rich \cuexcess, Mn-rich \mnexcess\ and near-stoichiometric \asexcess, were synthesized using traditional solid state synthesis routes. The elemental powders of Cu (99.9\% metals basis), Mn (99.98\% metals basis), and As (99.9999\% metals basis) were mixed in 1.18:0.82:1 ratio for the Cu-rich sample, 0.64:1.36:1 ratio in the Mn-rich sample, and in 1:0.964:1.036 ratio for the near-stoichiometric sample in an Ar-filled glovebox.
The mixed powders were vacuum sealed in quartz tubes and heated to 873~K in 10~h. The samples were held at 873~K for 6~h before heating to 1248~K at 1~K/min and held for 1~h. The Cu-rich and Mn-rich samples were cooled to 1173~K at 1~K/min and held for 1~h before furnace-cooling to room temperature. The near-stoichiometric sample was cooled slowly to 1023~K at 0.5~K/min and held for 1 h before cooling. Unlike the Cu-rich sample, mixed powders of the near-stoichiometric sample were transferred to an alumina crucible and the crucible was vacuum sealed inside a quartz tube in accordance with Uhlirova \textit{et al}. \cite{Uhlirova2019} The resulting ingots were black in color and lightly stuck to the tube or crucible. 

Variable-temperature synchrotron XRD measurements of the Cu-rich sample were taken using a nitrogen blower at beamline 11-BM of the Advanced Photon Source in Argonne National Laboratory \cite{wang_dedicated_2008}. Powder XRD measurements for the near-stoichiometric sample were performed in a Bruker D8 Advance in reflection geometry with a Cu source. Variable-temperature neutron powder diffraction (NPD) measurements for the Cu-rich and Mn-rich samples were carried out in POWGEN beamline at Spallation Neutron Source in Oak Ridge National Laboratory \cite{Huq:in5025,mason2006spallation}. 
Additional NPD for the Cu-rich sample were collected on the Wombat instrument at the Australian Center for Neutron Scattering (ACNS)\cite{STUDER20061013}. On Wombat, data were collected in a top loading cryostat in a vanadium can with copper heating blocks and a conduction arm between them at the top and the bottom of the sample, and an aluminium heat shield placed around both blocks and the sample. The NPD measurements were carried out at 4.6167(31)~\AA\ wavelength while cooling and at 2.41~\AA\ wavelength while heating. 
Data for the near-stoichiometric sample were collected on Echidna at ACNS \cite{Avdeev2018} at 2.43872(8)~\AA\ in a vanadium can at 4~K, 400~K, and 520~K in the same top-loading cryofurnace set-up as collected on Wombat. Rietveld analyses of XRD were performed using \textsc{GSAS-II} software \cite{Toby:aj5212}. The magnetic structure refinement of the NPD data with help from \textsc{k-Subgroupsmag} program \cite{Perez-Mato2015} in the Bilbao Crystallographic Server was carried out in \textsc{GSAS-II} \cite{Toby:aj5212}. The instrumental parameters for the Echidna data were obtained from the NIST SRM La$^{11}$B$_6$ 660b data and fixed in those sample refinements. More information on the refinement approach to the XRD and NPD data is provided in supplementary \cite{supplement}. The starting structures for CuMnAs and MnO refinements were taken from ICSD \# 423230 and 9864, respectively. The space group for the CuMnAs structure used in the refinements is $P4/nmm$.

Field cooling (FC) and zero field cooling (ZFC) curves with a field of 10~kOe for the samples were measured using a brass half-tube sample holder in a Quantum Design MPMS3 vibrating sample magnetometer. Powders of the samples weighing less than 10~mg were transferred to Al pans and differential scanning calorimetry (DSC) measurements were taken using a heat-cool-heat cycle between 93~K and 673~K at 10~K/min in a TA Instruments DSC 2500.

Scanning transmission electron microscopy (STEM) images and energy dispersive X-ray spectroscopy (EDS) elemental maps were obtained at room temperature using a Thermo Fisher Themis Z STEM operated at 300 kV with 25 mrad convergence angle.  Images were acquired with approximately 50 pA probe current and EDS data with approximately 500 pA.  EDS spectra were acquired over $2048 \times 2048$ probe positions.  EDS elemental maps were produced using the Velox software and smoothed with a moving average filter with a width of 41 pixels.

First-principles density functional theory (DFT) simulations were performed using the Vienna \emph{Ab-Initio} Simulation Package (\texttt{VASP})\cite{Kresse:1996,Kresse:1999}.
The Brillouin zone was sampled using an $18\,\times18\,\times10$ Monkhorst-Pack\cite{Monkhorst:1976} $\mathbf{k}$-point grid.
Kohn-Sham states were expanded into a plane wave basis with a kinetic-energy cutoff of 600\,eV.
These parameters allowed for converged calculations of the total energy within 0.6 meV per formula unit. 
The generalized-gradient approximation as parametrized by Perdew, Burke, and Ernzerhof\cite{Perdew:1997} (PBE) was used to describe the exchange and correlation contribution to the DFT Hamiltonian, in combination with an on-site Coulomb interaction, described using the DFT+$U$ approach of Dudarev \emph{et al.}\cite{Dudarev:1998}
$U_{\mathrm{eff}}$ values from the literature are adopted, as discussed in detail in Sec.\,\ref{sec:magstrc}.
We use a noncollinear description of magnetism including the effect of spin-orbit coupling \cite{Steiner:2016}.
Here we compare two cases:
First, we relax all atomic coordinates using the DFT ground state result for the magnetic structure, which shows the $Pm'mn$ (\# 59.407 in the Belov-Neronova-Smirnova (BNS) notation) magnetic space group with a magnetic moment on the Mn site only.
Next, to test if there is any tendency for local moments on Cu, we impose a constrained magnetic structure with a magnetic moment of 2.77\,$\mu_{B}$ on the Mn site and 0.50\,$\mu_{B}$ on the Cu site and relax again all atomic positions.

\section{Results and Discussion}





\subsection{Phase separation and heterogeneity}

\begin{figure}
\centering\includegraphics[width=\columnwidth]{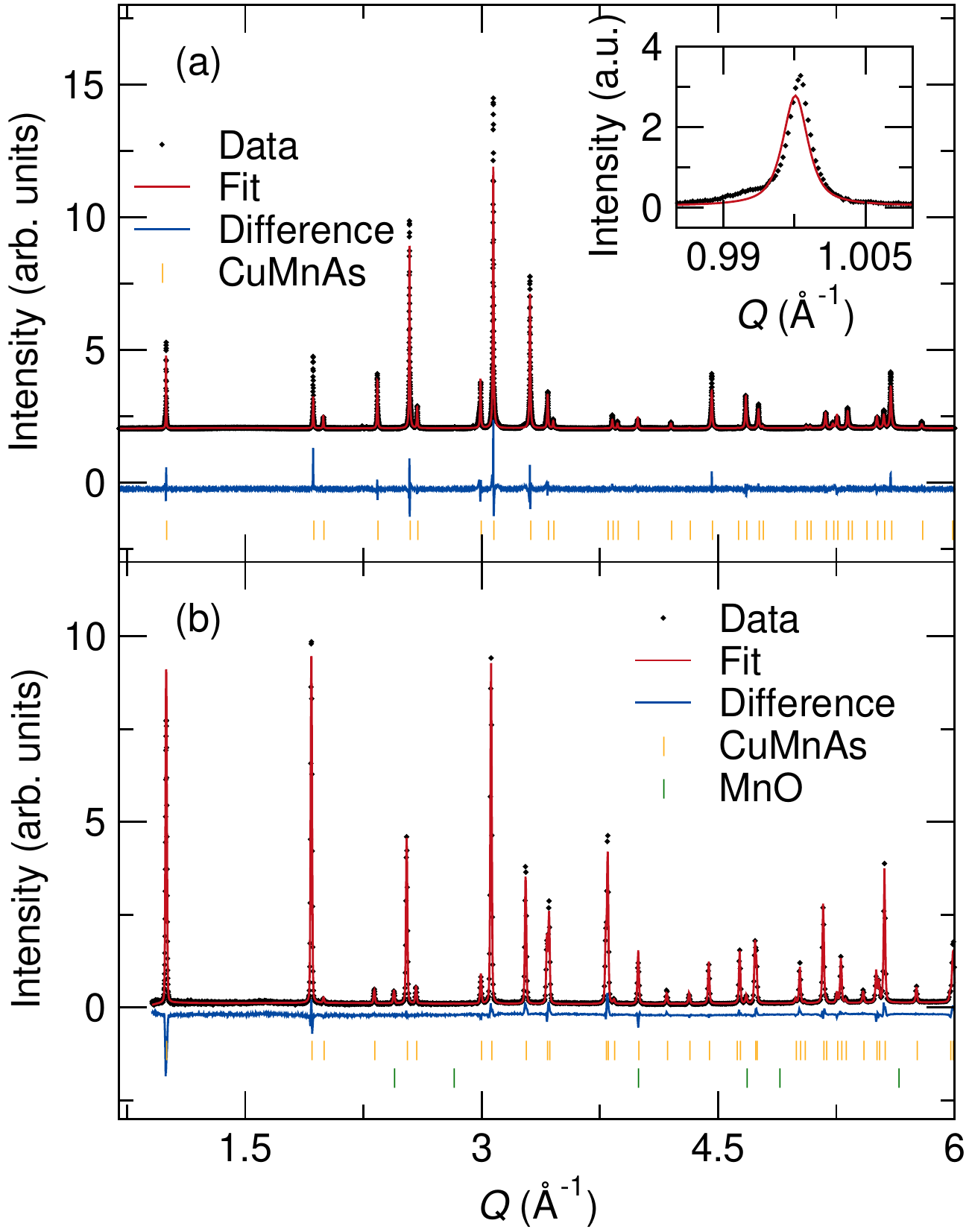} \\
\caption{\label{fig:Cu-rich_11BM_POWGEN}
 Rietveld fit to 298~K synchrotron XRD data of \cuexcess\ is shown in (a) and the fit to 500~K POWGEN NPD (above $T_N$, upon heating from 300~K) data is shown in (b). Asymmetry in the low-$Q$ peak inset (and the resulting poor fit) is evident due to the phase transition more visible in temperature-dependent measurements in Fig. \ref{fig:sampleA_transitions}. The material is single-phase at 500~K. 
} 
\end{figure}

\textbf{Cu-rich \cuexcess}: Fig. \ref{fig:Cu-rich_11BM_POWGEN}(a) shows the Rietveld fit to the synchrotron XRD data taken at 298~K. For laboratory x-ray diffraction data, the fit to tetragonal CuMnAs is satisfactory, but our high resolution data in Fig. \ref{fig:Cu-rich_11BM_POWGEN}(a) reveals a poor fit to the 001 peak, as visible in the inset. There is pervasive peak splitting from room temperature to 450~K, as we will discuss subsequently. 
X-rays cannot discriminate between Cu and Mn occupancies in CuMnAs due to their close electron counts. Neutrons, on the other hand, give distinct scattering lengths for Cu and Mn (7.72 and $-$3.73 fm, respectively). From fits to 500~K NPD measurements in Fig. \ref{fig:Cu-rich_11BM_POWGEN}(b), the peak splitting is not apparent and the refined Cu:Mn stoichiometry was obtained as 1.186(3):0.814(3), which matches the nominal synthesis stoichiometry. In the refinement, Cu was allowed to partially occupy Mn sites and the total occupancy was constrained to be 1 at the Mn site. When Mn was also allowed to partially occupy Cu sites, the refinement yielded negligible values for Mn occupancy. This proves that, on average, excess Cu substitutes into Mn sites in \cuexcess.

\begin{figure}
\centering\includegraphics[width=\columnwidth]{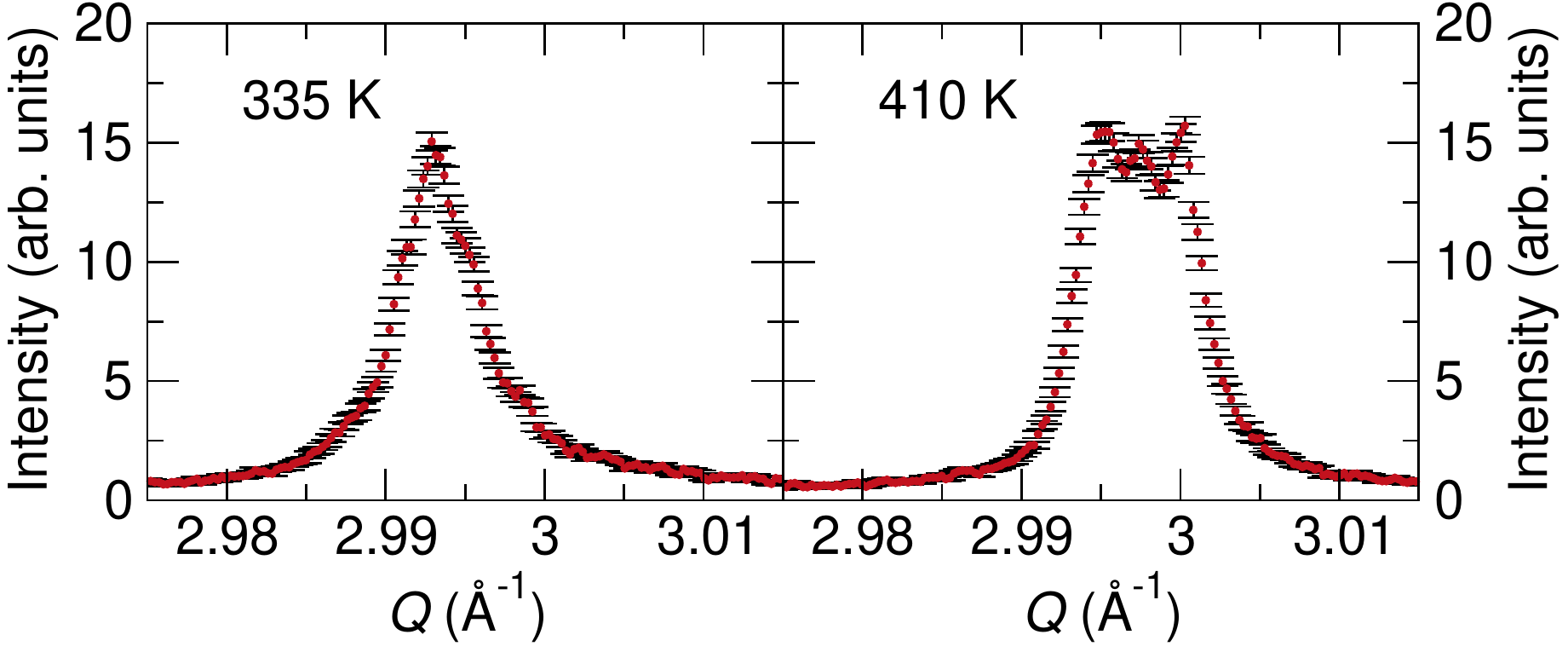} \\
\caption{\label{fig:peak_splitting}
The 003 synchrotron XRD peak is shown at 335~K and 410~K for the same \cuexcess\ sample as the temperature map in Fig. \ref{fig:sampleA_transitions}(a). At least three peaks are apparent at 410~K.
} 
\end{figure}

\begin{figure}
\centering\includegraphics[width=\columnwidth]{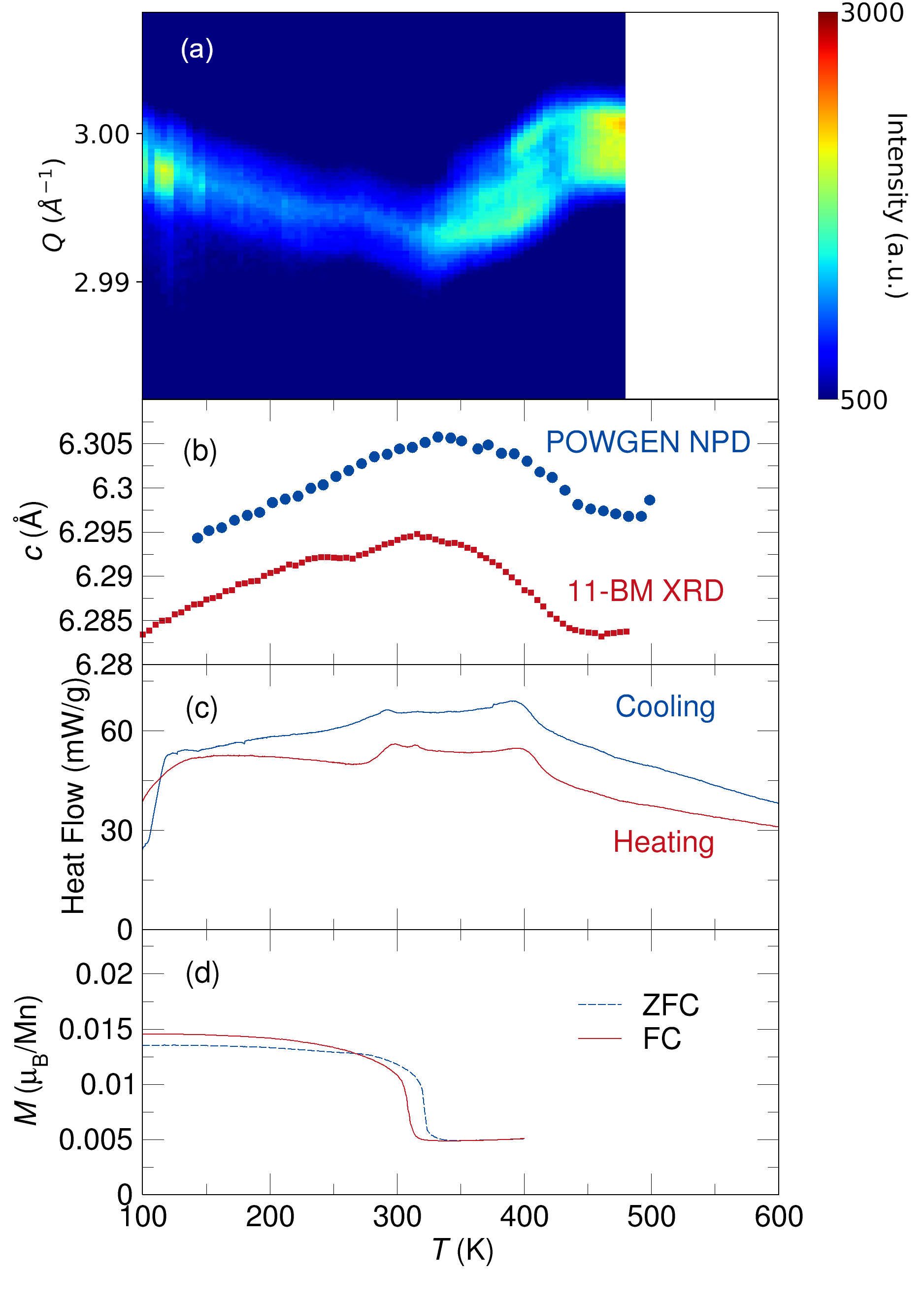} \\
\caption{\label{fig:sampleA_transitions}
For the Cu-rich sample \cuexcess, the in-situ synchrotron XRD measurements (collected on heating) have sufficient resolution to observe splitting in the higher-order 003 peak, which is shown in (a). (b) shows the change in $c$ lattice parameter across temperature as determined from a single-peak fit to the 001 reflections in synchrotron XRD and POWGEN NPD measurements. The DSC heating (inverted) and cooling curves are shown in (c) and the field cooling and zero field cooling curves are shown in (d). More than two phases are present in the intermediate temperature range from 300 to 400~K.
} 
\end{figure}

\begin{figure}
\centering
\includegraphics[width=0.8\columnwidth]{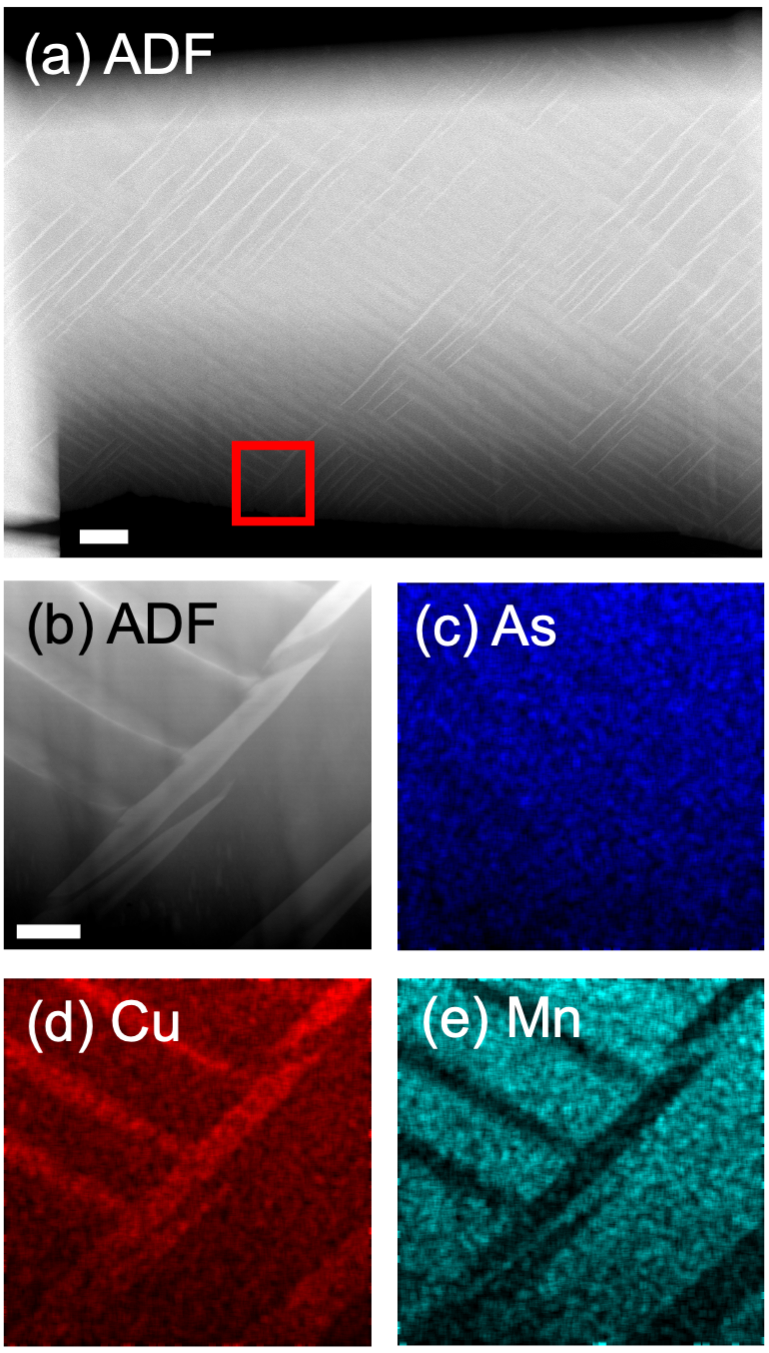} \\
\caption{\label{fig:TEM_imaging}
For the Cu-rich \cuexcess\ sample, STEM micrographs of a FIB liftout reveal a tweed-like pattern in ADF imaging. The boxed area is shown in panels (b-e) for ADF, As, Cu, and Mn EDS maps. Scale bars are 500~nm in (a) and 100~nm in (b-e).
} 
\end{figure}

While there were subtle peak splittings in the synchrotron XRD data shown in Fig. \ref{fig:Cu-rich_11BM_POWGEN}(a), a close look at the 003 XRD peaks upon heating shows more complex and pervasive changes. The splitting of the 003 XRD peak into three peaks is shown in Fig. \ref{fig:peak_splitting}, and as a contour plot versus temperature in Fig. \ref{fig:sampleA_transitions}(a). 
We later show that the peak splitting occurs due to chemical inhomogeneity and phase separation as evidenced in the TEM images shown in Fig. \ref{fig:TEM_imaging}. The peak splitting behavior is not resolvable in the POWGEN NPD measurements. However, if an average $c$ lattice parameter is estimated as a peak fit to the 001 reflection for all measurements, consistent trends appear from POWGEN neutron measurements and 11-BM synchrotron measurements as shown in Fig. \ref{fig:sampleA_transitions}(b). The discrepancy in the $c$ lattice parameters between the two measurements can be attributed to peak shift effects such as sample displacement that is not taken into account during peak fitting. Sequential peak fits to the 001 reflection of the Wombat neutron data in Fig. S1 \cite{supplement} shows a lack of change in the slope of the $c$ lattice parameter below 300~K.

Fig. \ref{fig:sampleA_transitions}(c) shows the results of DSC measurements for the \cuexcess\ sample. We observe two kinks at around 300~K and 420~K respectively. The transition at 420~K corresponds to the T$_N$ of the sample. This is also confirmed from studies by Uhlirova \emph{et al}.\cite{Uhlirova2019} where increasing the Cu substitution at Mn sites decreases the T$_N$ considerably from its maximum value of around 520~K. The transition at 300~K has a clear signature in magnetic susceptibility shown in Fig. \ref{fig:sampleA_transitions}(d). There is an increase in the net moment below 300~K indicating a possible transition from an AFM to a weak FM phase. As mentioned earlier, the ferromagnetic transition has been reported in previous studies,  \cite{Nateprov2011,Uhlirova2019} although it was tentatively attributed to the presence of a possible MnAs impurity. This explanation is unlikely since we do not observe any MnAs impurity in the synchrotron XRD data of the Cu-rich sample as shown in Fig. \ref{fig:Cu-rich_11BM_POWGEN}(a), and the presence of a phase transition in the majority CuMnAs phase is clear from in-situ diffraction data and calorimetry in Fig. \ref{fig:sampleA_transitions}(a,b,c). Similar transitions are also observed in the DSC and SQUID measurements of the near-stoichiometric sample, as we discuss subsequently.
Such changes in lattice parameters around $T_N$ are not observed in Fe$_2$As, which has the same structure type as CuMnAs, as shown in Fig. S2 \cite{supplement}. 


Since diffraction measurements did not permit refinement of Cu/Mn occupancies in both phases simultaneously, a focused ion beam (FIB) cross-section of the Cu-rich \cuexcess\ sample was examined via STEM. Fig. \ref{fig:TEM_imaging}(a) shows the microstructure and STEM-EDS elemental mapping of a polished surface in the Cu-rich sample. Aligned stripes of two distinct phases are clearly present in the annular dark field (ADF) STEM image. EDS Elemental mapping could accurately measure the Cu content due to the background from the Cu grid, but the Mn:As ratios in the two samples were measured to be approximately 0.2 and 0.8 in the bright and dark regions, respectively. This is likely an underestimation since the nominal and neutron-refined Mn:As ratio is 0.82, but it is clear that the chemical separation is pervasive.  More ADF-STEM images of the Cu-rich \cuexcess\ sample are shown in Fig. S3 \cite{supplement}. Fig. S3(a) \cite{supplement} shows heterogeneity even outside the region containing the tweed-like patterns shown in Fig. \ref{fig:TEM_imaging}(a).

The interdependence of phase separation and magnetism hint at strong coupling, reminiscent of martensitic shape memory alloys.  This type of phase separation of topologically connected phases with subtle chemical phase separation has also been observed   in intermetallics and perovskite and spinel oxides.\cite{le_bouar_origin_1998,guiton_nano-chessboard_2007,park_highly_2008} Ni and Khachaturyan presented a detailed model showing how a pseudospinodal transformation from a high-symmetry phase to lower-symmetry components can result in  tweed-like structures seen in metal alloys and oxide ceramics. \cite{Ni2009}
The precise reason for this phase separation is not known. 
The metallicity and lack of local moments on Cu in CuMnAs argues against a Jahn-Teller-driven effect, further evidenced by the fact that the phase separation seems to disappear upon cooling below 300 K (see Fig. \ref{fig:sampleA_transitions}(a,b). Rather, the onset of magnetic ordering must drive pervasive lattice distortion that merits further investigation into magnetostrictive effects in Cu$_2$Sb-type structures, including CuMnAs.

\begin{figure}
\centering\includegraphics[width=\columnwidth]{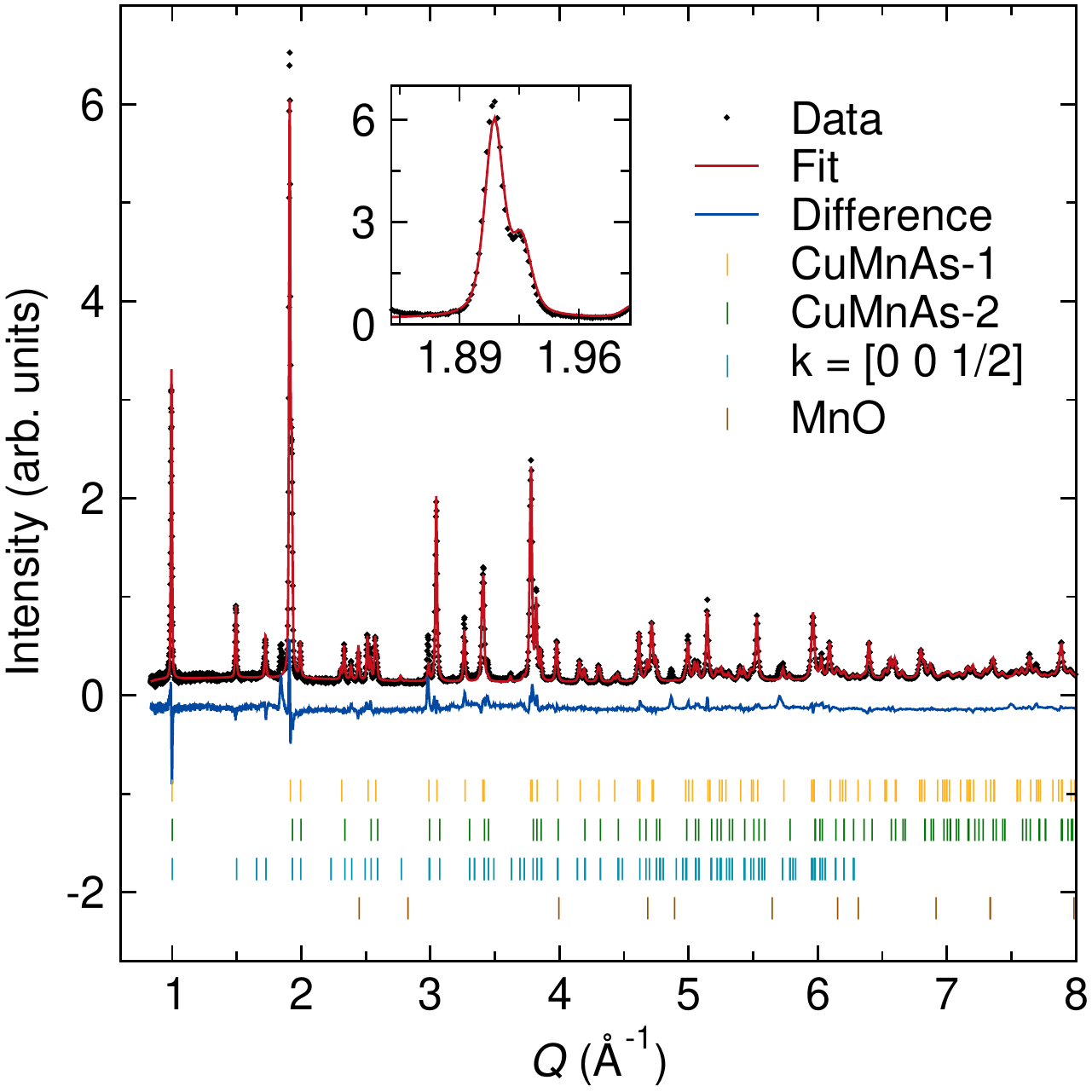} \\
\caption{\label{fig:Mn-rich_500K_NPD}
Rietveld fit to the POWGEN NPD data of the Mn-rich sample \mnexcess\ at 500~K shows the presence of two closely-related tetragonal CuMnAs phases. The inset shows the split 101 peak that indicates a clear phase separation.
One of the phases displays a Mn$_2$As-like $k=[0 0 \frac{1}{2}]$ magnetic peak (as opposed to the $k=0$ magnetic structure of near-stoichiometric CuMnAs). The peak at $Q = 1.9$~\AA\ could not be assigned, and appeared irreversibly upon initial heating.
} 
\end{figure}
\begin{figure}
\centering\includegraphics[width=\columnwidth]{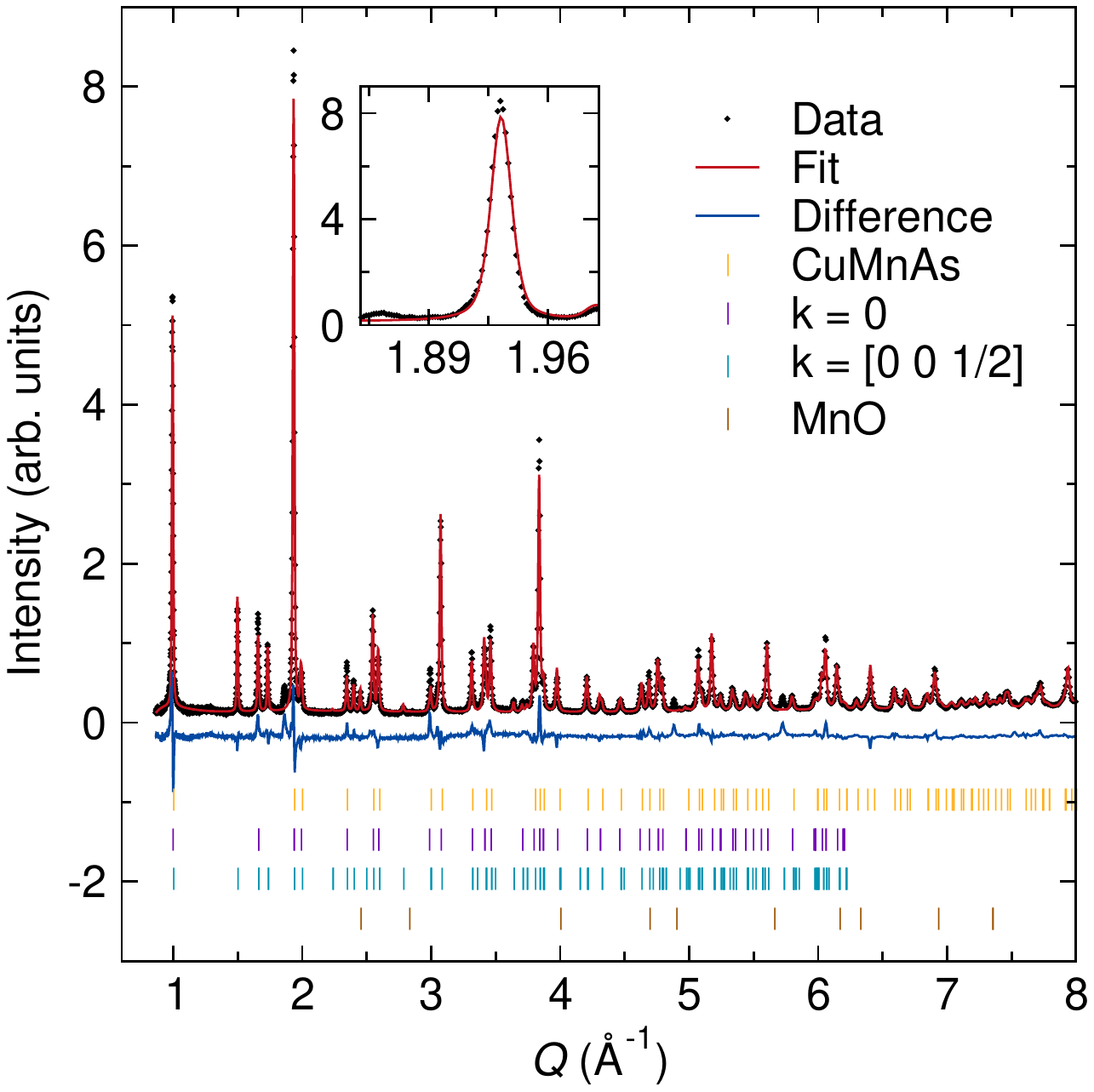} \\
\caption{\label{fig:Mn-rich_300K_NPD}
Rietveld fit to the POWGEN NPD data of the Mn-rich \mnexcess\ sample at 300~K. The inset figure shows that unlike at 500~K, the 101 peaks does not show splitting, so there is no evidence for phase separation from structural peaks alone. However, magnetic peaks are present for both the $k=0$ (CuMnAs-like) and $k= 0 0 \frac{1}{2}$ (Mn$_2$As-like) magnetic ordering.
} 
\end{figure}

\textbf{Mn-rich \mnexcess}: Similar phase separation is observed in the Mn-rich composition, even at 500~K as observed by a clear splitting of the 101 peak in NPD data shown in Fig. \ref{fig:Mn-rich_500K_NPD}. 
At 300~K, NPD data in Fig. \ref{fig:Mn-rich_300K_NPD} do not show split peaks and the material is apparently a single structural phase.
This phase separation behavior may mimic the appearance of the Cu-rich composition, with the multiple-phase region in the Mn-rich sample (with more magnetic moment) shifted to higher temperatures than the Cu-rich compositions. 
Regardless of the lack of splitting of structural peaks at room temperature, the magnetic peak contributions in NPD patterns cannot be fit with a single $k$-vector: the presence of magnetic peaks corresponding to both $k=0$ and $k=[00\frac{1}{2}]$ ordering at 300~K in Fig. \ref{fig:Mn-rich_300K_NPD} reveals the existence of both the established tetragonal CuMnAs $k=0$ magnetic structure (alternating left/right moments along the c-axis) and an Mn$_2$As-like $k=[00\frac{1}{2}]$ magnetic ordering.
This apparent coexistence may appear due to subtle variations in local Cu/Mn concentrations, as has been seen before in two-dimensional magnetic materials that are apparently structurally phase-pure \cite{Bhutani2020}. 

\begin{figure}
\centering\includegraphics[width=\columnwidth]{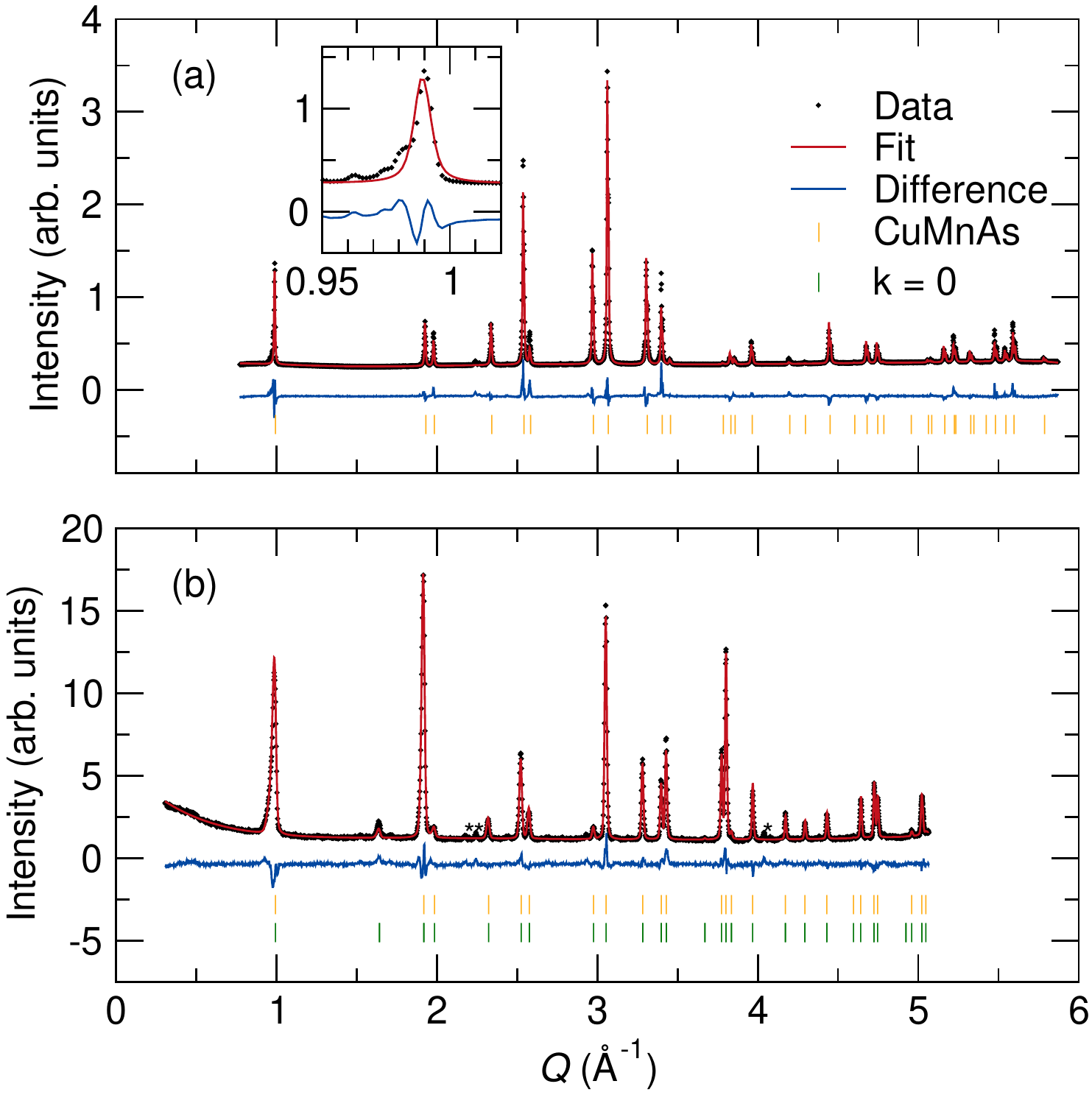} \\
\caption{\label{fig:XRD_NPD_near_stoichiometric}
Rietveld fits for the near-stoichiometric sample \asexcess\ to the room temperature XRD data in (a) roughly fit the data, but some small amount of inhomogeneity is evident in the 001 peak, magnified in the inset. This asymmetry is not evident in the high-temperature 520 K Echidna NPD data in (b). 
In (b), the peak at around $Q = 1.65$~\AA$^{-1}$ is a magnetic peak and the small peaks at around $Q = 2.25$~\AA$^{-1}$ and 4.1~\AA$^{-1}$ (marked by *) are impurity peaks of unknown origin.
} 
\end{figure}

\textbf{Near-stoichiometric \asexcess}: In case of the near-stoichiometric composition, good fits to room temperature synchrotron XRD and 520~K NPD measurements were obtained by fixing the Cu and Mn occupancies to be stoichiometric, as seen in Fig. \ref{fig:XRD_NPD_near_stoichiometric}. While some anti-site mixing of Cu and Mn is possible, there was no indication of peak splitting in the 300~K and 520~K data. However, DSC and magnetic susceptibility measurements in Fig.\ S4 \cite{supplement} show that two transitions indeed exist within this temperature range, at 315 and 490~K. Therefore, even the near-stoichiometric sample without Cu/Mn excess or mixing likely shows the same phase separation over a similar temperature range. 

The T$_N$ of the near-stoichiometric sample was observed to be around 490~K and the AFM-FM transformation was confirmed at 315~K which is consistent with the values reported in Uhlirova \emph{et al}. \cite{Uhlirova2019}

\subsection{\label{sec:magstrc}Magnetic structure confirmation and magnetoelastic effects}

A structurally-forbidden 100 magnetic peak appears at $Q$~=~1.65~\AA$^{-1}$ upon cooling Cu-rich \cuexcess\ and near-stoichiometric \asexcess\ below T$_N$. This peak is most visible in the ECHIDNA NPD data at 4~K as shown for near-stoichiometric \asexcess\ in Fig. S5\cite{supplement} and POWGEN 300~K NPD data for \cuexcess\ in Fig. S6\cite{supplement}.
This corresponds to the expected $k = 0$ magnetic ordering vector. With $P4/nmm$ as the parent space group and a $k = 0$ propagation vector, there are 12 $k$-maximal space groups, of which 4 are in the orthorhombic crystal system as shown in Table S1\cite{supplement}. None of the 8 magnetic subgroups in the tetragonal crystal system allow Mn moment components in the $ab$ plane. Of the 4 $k$-maximal magnetic subgroups in the orthorhombic crystal system, Mn moments are ordered ferromagnetically in two of them. Of the remaining two models, a slightly better fit is obtained for the model with a $Pm'mn$ (\#59.407) magnetic space group (MSG) shown in Fig. S5 \cite{supplement} which is also reported in a previous study of CuMnAs films\cite{Wadley2015,supplement}. The refined magnetic moment on Mn is 3.73(3) $\mu_B$.

The small uncompensated moment that arises in the Cu-rich sample at around 300~K and at 315~K in the near-stoichiometric sample requires further investigation. It is clearly an intrinsic effect since it is accompanied by a change in the lattice parameters (Fig.\ \ref{fig:sampleA_transitions}(b)). The Rietveld fit to the NPD data of the near-stoichiometric sample at 4~K using $Pm'mn$ MSG is shown in Fig. S5 \cite{supplement}. The fit is satisfactory and there is no significant improvement of fit using any of the magnetic subgroups of $Pm'mn$. We assume that the small moment arises from a canting of spins at a weak-ferromagnetic transition, which cannot be resolved from NPD measurements. Single-crystal susceptibility and neutron diffraction measurements could be performed if the factors affecting phase separation and crystal growth in CuMnAs can be controlled. 

We use density functional theory (DFT) to describe the ground state lattice and magnetic structure of tetragonal CuMnAs.
Previous studies typically use a Hubbard correction within the DFT+$U$ method \cite{Anisimov:1991,Anisimov:1997} to investigate tetragonal CuMnAs, using an on-site Coulomb term $U$ and site exchange term $J$.
Here a simplified approach is used with an effective on-site Coulomb interaction $U_\mathrm{eff}=U-J$ for which values of 1.7\,eV\cite{Veis2018}, 1.92\,eV\cite{Zhuravlev2018}, 3.0\,eV\cite{Xu2020}, and 4.1\,eV\cite{Wadley2013} were reported.
Here we wish to examine how the magnetic ordering and moment magnitude affects the lattice, and whether there is any tendency for local moments on Cu atoms, for different values of $U_{\mathrm{eff}}$.

\begin{figure}[h]
\centering\includegraphics[width=0.9\columnwidth]{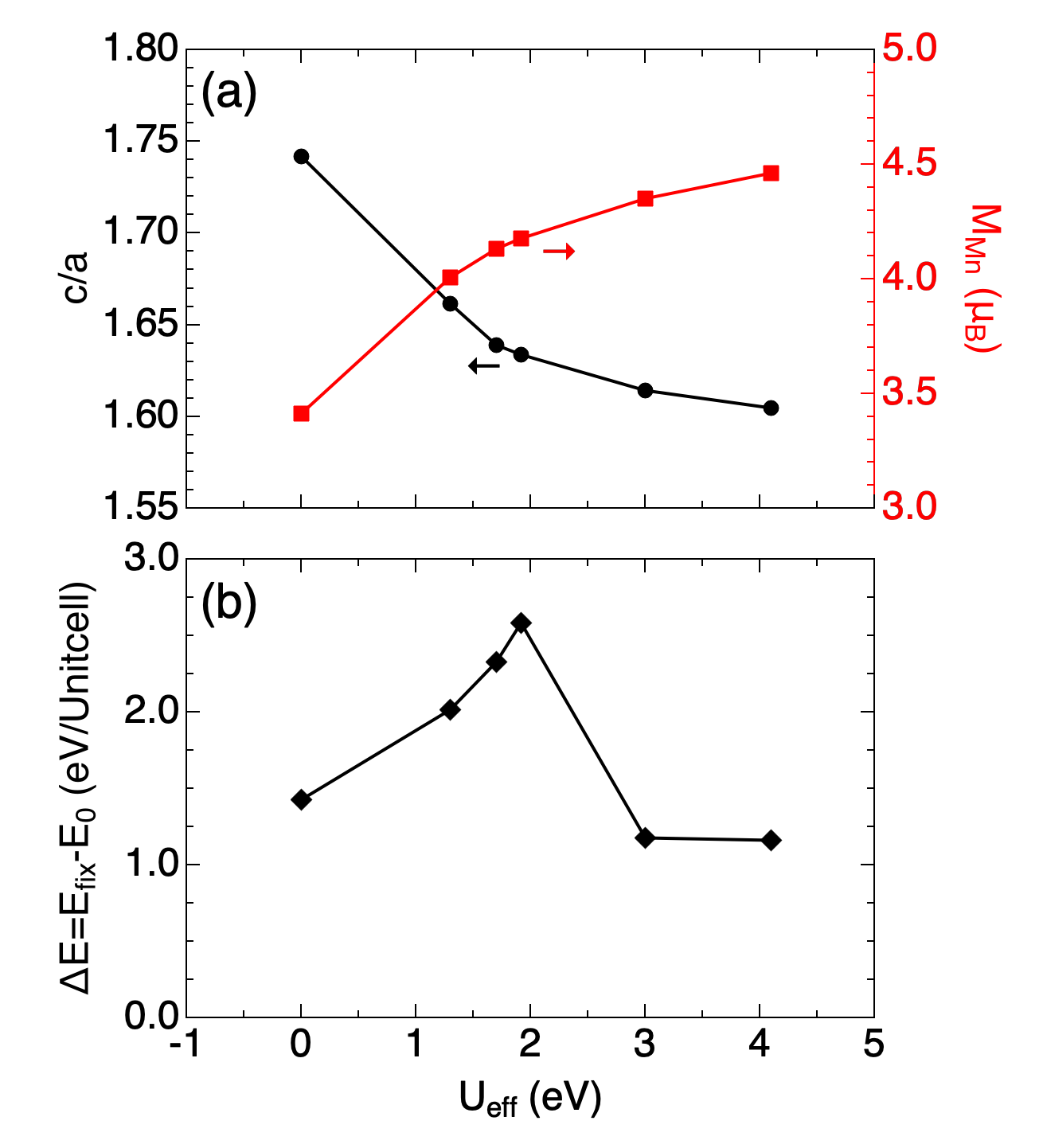} \\
\caption{\label{fig:DFTU}
(a) The lattice parameter ratio ($c/a$, black circles) and the magnitude of the magnetic moment at Mn atom (M$_{\mathrm{Mn}}$, red squares), and (b) the energy difference between the ground state and the constrained magnetic structure with a magnetic moment at Cu atom ($\Delta$E, black diamond) as a function of the effective on-site Coulomb interaction U$_{\mathrm{eff}}$ value in DFT calculations. The lines through the points are guides to the eye.
} 
\end{figure}

With increasing $U_{\mathrm{eff}}$, the ratio of the lattice parameters $c/a$ decreases from 1.742 to 1.605. 
The room-temperature experimental $c/a$ ratios for the three compositions in our study span a much narrower range between 1.665 and 1.700. 
The calculated magnetic moment on Mn sites increases with $U_{\mathrm{eff}}$, increasing from 3.412\,$\mu_{B}$ to 4.462\,$\mu_{B}$ (see Fig. \ref{fig:DFTU}(a)). 
Taking the value of $c/a$ for the near-stoichiometric \asexcess\ sample to prescribe $U_{\mathrm{eff}}$ would correspond to a calculated Mn magnetic moment of 4.0 $\mu_b$, which is in rough agreement with the refined value of 3.73(3) $\mu_B$ at 4 K.

\begin{figure}[h]
\centering\includegraphics[width=0.8\columnwidth]{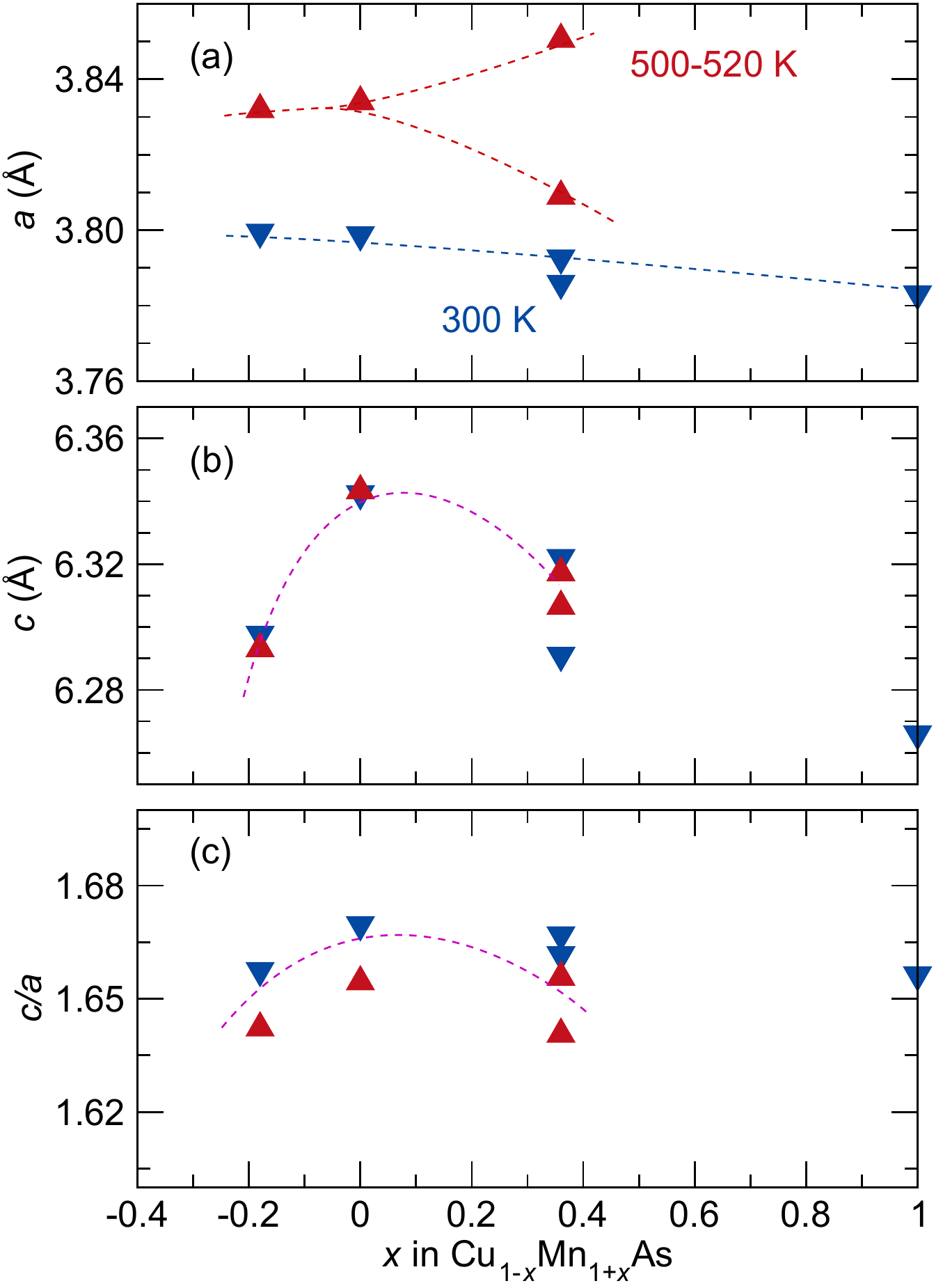} \\
\caption{\label{fig:lattice_params}
The lattice parameters across various compositions  obtained using NPD refinements to $T=300$ and 500~K data. $T=300$~K data for the Cu-excess and the near-stoichiometric sample are taken from x-ray diffraction measurements. Compositions are nominal. Values for Mn$_2$As are from Nuss, et al.\cite{nuss2006geometric} It is clear in (a) that an opposing trend in composition versus $a$ appears above and below $T_N$, but this trend must be broken as $x$ increases toward Mn$_2$As, and the \mnexcess\ composition has separated into two phases. In (b), the relative splittings in $c$ are less variable with temperature but exhibit a maximum around the equiatomic CuMnAs composition.  The $c/a$ ratios in $c$ are dominated by the trends in $c$. The dashed lines are used as a guide for observing trends.
} 
\end{figure}

The $U_{\mathrm{eff}}$-dependence of $c/a$ and Mn magnetic moment in Fig. \ref{fig:DFTU}(a) show opposing trends, indicating that increased Mn magnetic moments (or increasing exchange interactions between these sites, as embodied in $U_{\mathrm{eff}}$) within the $ab$ planes may lead to a flattening of the cell, as $c$ decreases and $a$ increases. Magnetic interactions also must contribute to the lattice parameters observed across the compositions series, where the room-temperature in-plane $a$ decreases with increasing Mn content across the entire compositional range, seen in Fig. \ref{fig:lattice_params}. However, the $c$ axis parameters at room temperature exhibit a maximum around the equiatomic CuMnAs composition. Deviating from this stoichiometry leads to regions where strong Mn exchange interactions are unsatisfied (Cu-rich) or interact with additional Mn in the nearest-neighbor plane (Mn-rich). The pervasive phase separations in these systems may be driven by nanometer-scale domains where the local magnetic ordering leads to $k=0$ or $k=00\frac{1}{2}$-type ordering (as an example), which each have distinct magnetostrictive distortions to differing $c/a$ ratios around $T_N$. 
Controlling the formation of these domains and their sizes will require investigation of high-temperature processing, where the solidus behavior and tendency to form orthorhombic CuMnAs is still unknown. The strong compositional dependence of the stripes in Fig.\ \ref{fig:TEM_imaging} further indicate that Cu and Mn species may be mobile at the N\'{e}el temperature and weak ferromagnetic transition temperatures in CuMnAs. 

Finally, DFT simulations of the ground-state magnetic structure of tetragonal CuMnAs do not show significant magnetic moments on Cu atoms, regardless of the on-site Coulomb interaction values studied here.
In addition, our simulations show that the magnetic structure with non-negligible Cu magnetic moments in the constrained magnetic configurations is energetically unfavorable, regardless of the $U_\mathrm{eff}$ values studied here (see Fig. \ref{fig:DFTU}(b)).
We also found that, in the ground state structure, the magnetization density is centrally distributed near Mn atoms and there is no population of magnetization density near Cu atoms (see Fig.\ S7 \cite{supplement}).
Based on Bader charge analysis\cite{Tang:2009}, the charge on Cu atoms is close to neutral ($+0.1$) which is far from the $3d^9$ Cu$^{2+}$ state  observed in cuprates\cite{DeLuca:2010}, and CuMnAs only hosts local moments on Mn.
Lastly, we have implemented a total energy comparison among different types of magnetic structures from DFT simulations. The refined collinear antiferromagnetic structure has the lowest total energy among three antiferromagnetic structures and one ferromagnetic structure, confirming the agreement between the experiment and DFT for the ground state magnetic configuration (see Fig.\ S8 \cite{supplement}).
A remaining question is whether magnetic ordering drives the phase separation or vice versa. Computational insight on this point may be challenging since the lattice response to compositional change is much stronger than the response to crossing $T_N$ (Fig. \ref{fig:lattice_params}), yet phase separation into multiple compositions is pervasive in the Cu-Mn-As system.

\section{Conclusions}

The complex interplay between local stoichiometry, competing magnetically-ordered states, and magnetoelastically-driven phase separation must be considered when analyzing the results of any magnetic domain-related spintronic application in CuMnAs, whether in the bulk or in thin films, where substrate strain adds a further consideration. 
Pervasive phase separation is observed in three samples, Cu-rich \cuexcess, Mn-rich \mnexcess\ and \asexcess\ which is near-stoichiometric. 
Diffraction, magnetometry, and calorimetry measurements all point to the presence of two intrinsic magnetic transitions corresponding to T$_N$ and an AFM to weak ferromagnetic transition. Lattice constants extracted from synchrotron XRD and NPD data indicate a coupling between the structural and magnetic order. A clear trend can be drawn in the predicted $c/a$ ratio as affected by $U_{\mathrm{eff}}$, which implies that the phase separation likely arises from disordered regions of Cu or Mn clustering that are rearranged at the magnetic ordering temperatures. A clear understanding of the high-temperature behavior of these phases will aid the synthesis of single-crystal or single-phase materials, and the full picture of the phase equilibria across the Cu/Mn ratios and immiscibility behavior. Any time there is a study on the N\'eel order spin orbit torque with tetragonal Cu-Mn-As, we have to consider the effect of a Mn$_2$As based magnetic ordering as well. In same cases, such as in \mnexcess, this phase is magnetic. The weak FM transition in these materials could provide an opportunity to image magnetic domains.

\begin{acknowledgments}
This work was undertaken as part of the Illinois Materials Research Science and Engineering Center, supported by the National Science Foundation MRSEC program under NSF Award No.\ DMR-1720633.
The characterization was carried out in part in the Materials Research Laboratory Central Research Facilities, University of Illinois.
This research used resources of the Spallation Neutron Source, a DOE Office of Science User Facility operated by Oak Ridge National Laboratory, and the Advanced Photon Source, a DOE Office of Science User Facility operated for the DOE Office of Science by Argonne National Laboratory under Contract No.\ DEAC02-06CH11357.
The NPD measurements in Wombat and Echidna beamlines were carried out under the proposals number DB8036 and MI1931, respectively. We thank Matthias Frontzek for assistance with NPD collection in POWGEN beamline at the Spallation Neutron Source.
This work made use of the Illinois Campus Cluster, a computing resource that is operated by the Illinois Campus Cluster Program (ICCP) in conjunction with the National Center for Supercomputing Applications (NCSA) and which is supported by funds from the University of Illinois at Urbana-Champaign.
\end{acknowledgments}
\bibliography{CuMnAs_ordering}

\end{document}



\begin{center}
\Large 
\textbf{High-resolution diffraction reveals magnetoelastic coupling and coherent phase separation in tetragonal CuMnAs}\\
\vspace{1em}
Supplementary Material\\
\vspace{1em}
\normalsize
Manohar H. Karigerasi, Kisung Kang, Jeffrey Huang, Vanessa K. Peterson, Kirrily C. Rule, Andrew J. Studer, Andr\'e Schleife, Pinshane Y. Huang, Daniel P. Shoemaker

\vspace{2em}
\end{center}

\begin{center}

\begin{figure}[h]
\centering\includegraphics[width=0.7\columnwidth]{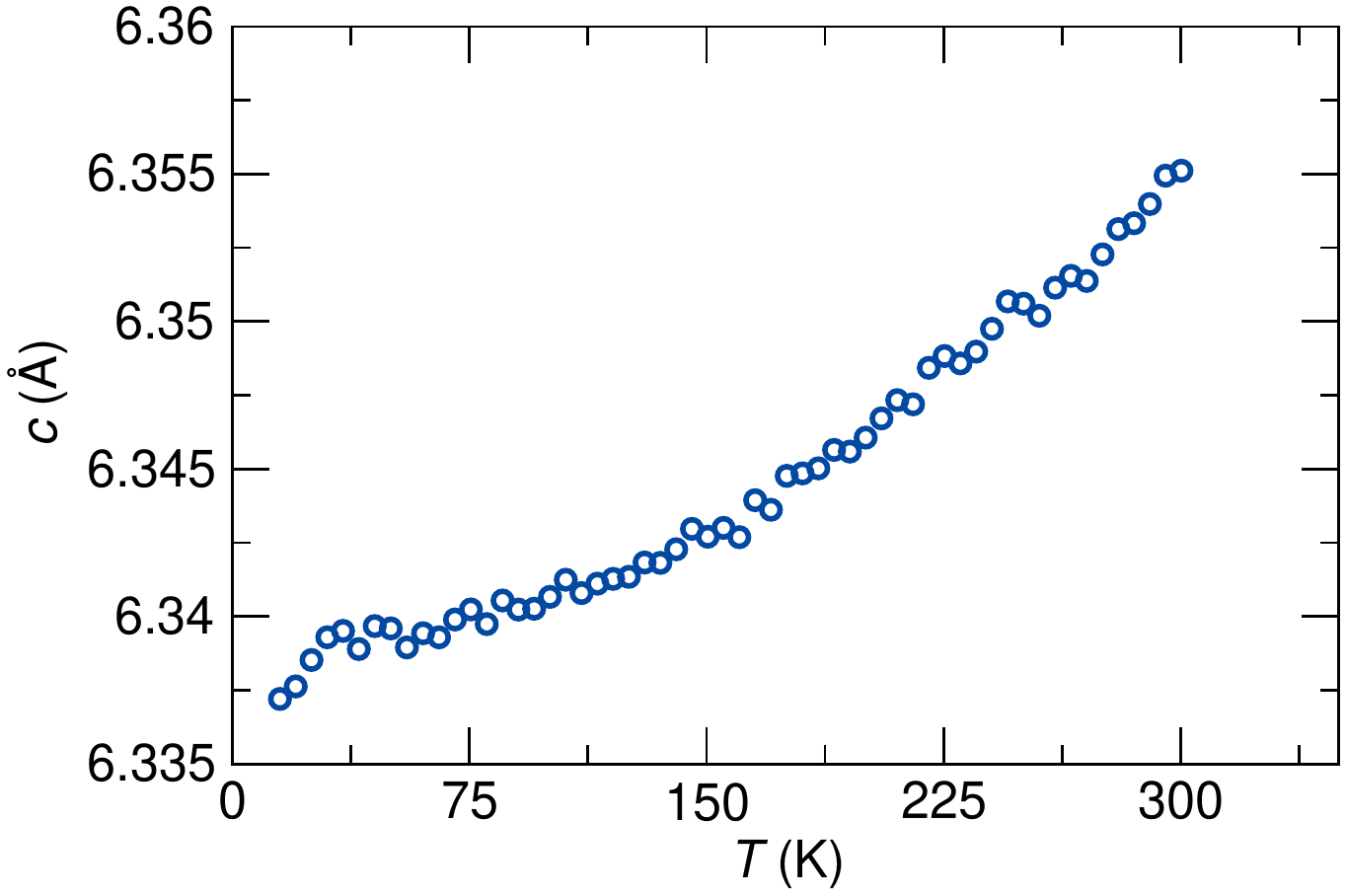} \\
\caption{\label{fig:CuMnAs_Wombat_c}
$c$ lattice parameter as a function of temperature of \cuexcess\ extracted from a sequential peak fitting to the 001 peak in the Wombat NPD data. The values do not incorporate shifts due to sample displacements or other instrumental parameters.
} 
\end{figure}

\begin{figure}[h]
\centering\includegraphics[width=0.7\columnwidth]{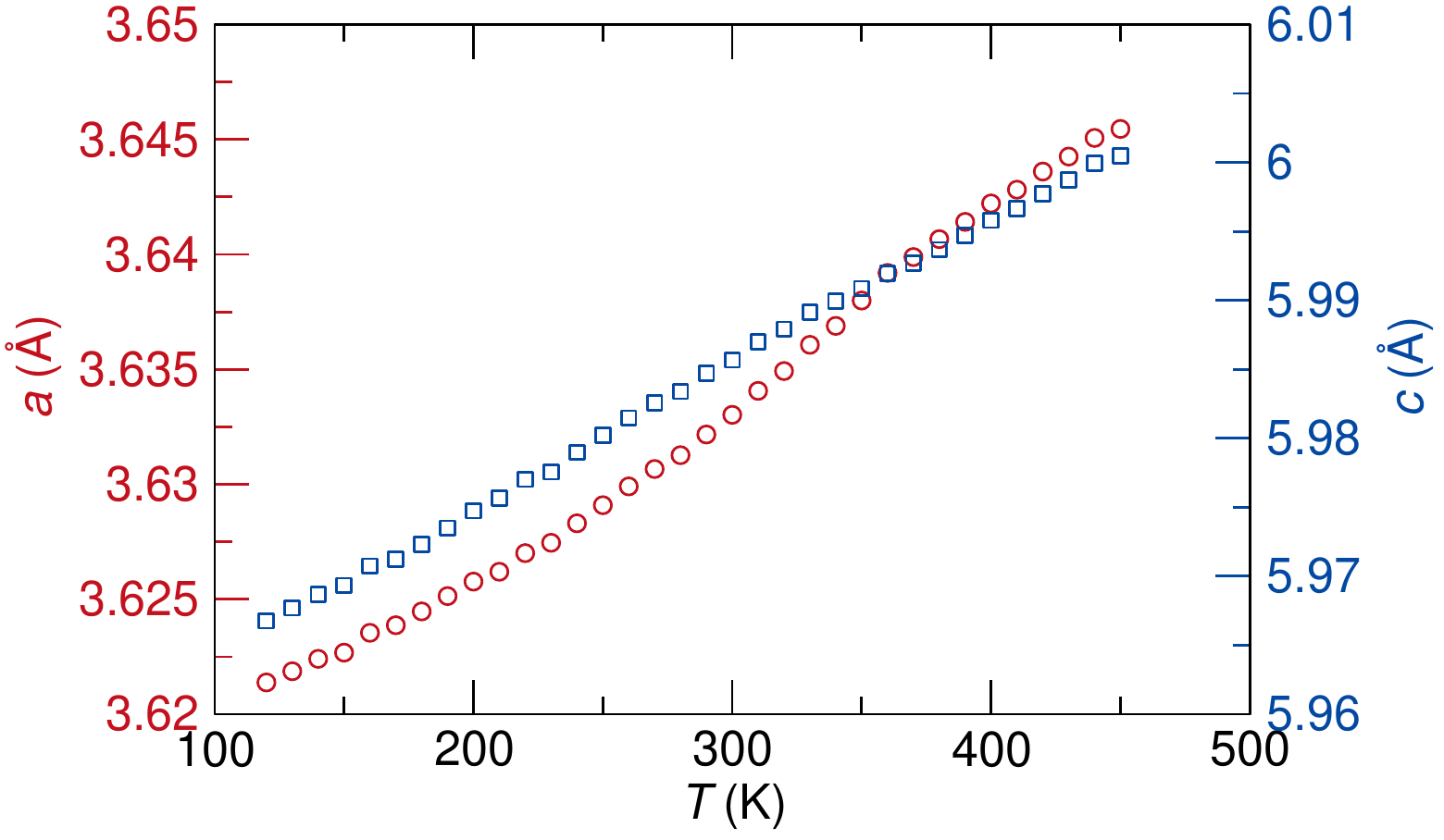} \\
\caption{\label{fig:Fe2As}
$a$ and $c$ lattice parameters of Fe$_2$As across a range of temperatures obtained from Rietveld refinement of synchrotron XRD data. More information on the synthesis and XRD measurements of Fe$_2$As can be found in Karigerasi \emph{et al}. (2020).
} 
\end{figure}

\begin{figure}[h]
\centering\includegraphics[width=\columnwidth]{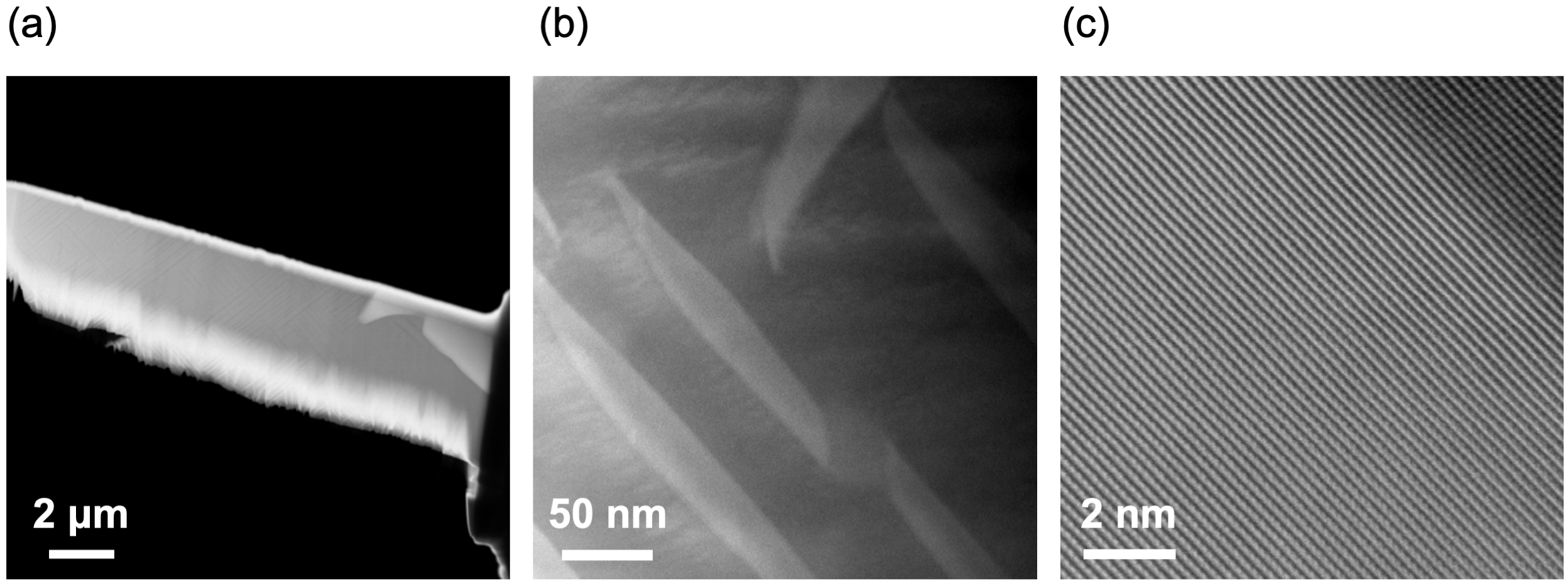} \\
\caption{\label{fig:Cu-rich_phase_separation}
ADF-STEM images of the Cu-rich sample at different magnifications. (a) shows the zoomed-out view of the TEM sample where we can see two different regions of contrast. (b) shows the zoom-in view of the dark region in (a). (c) shows the crystal lattice mostly of the lighter Cu-rich phase. 
} 
\end{figure}


\begin{figure}[h]
\centering\includegraphics[width=0.9\columnwidth]{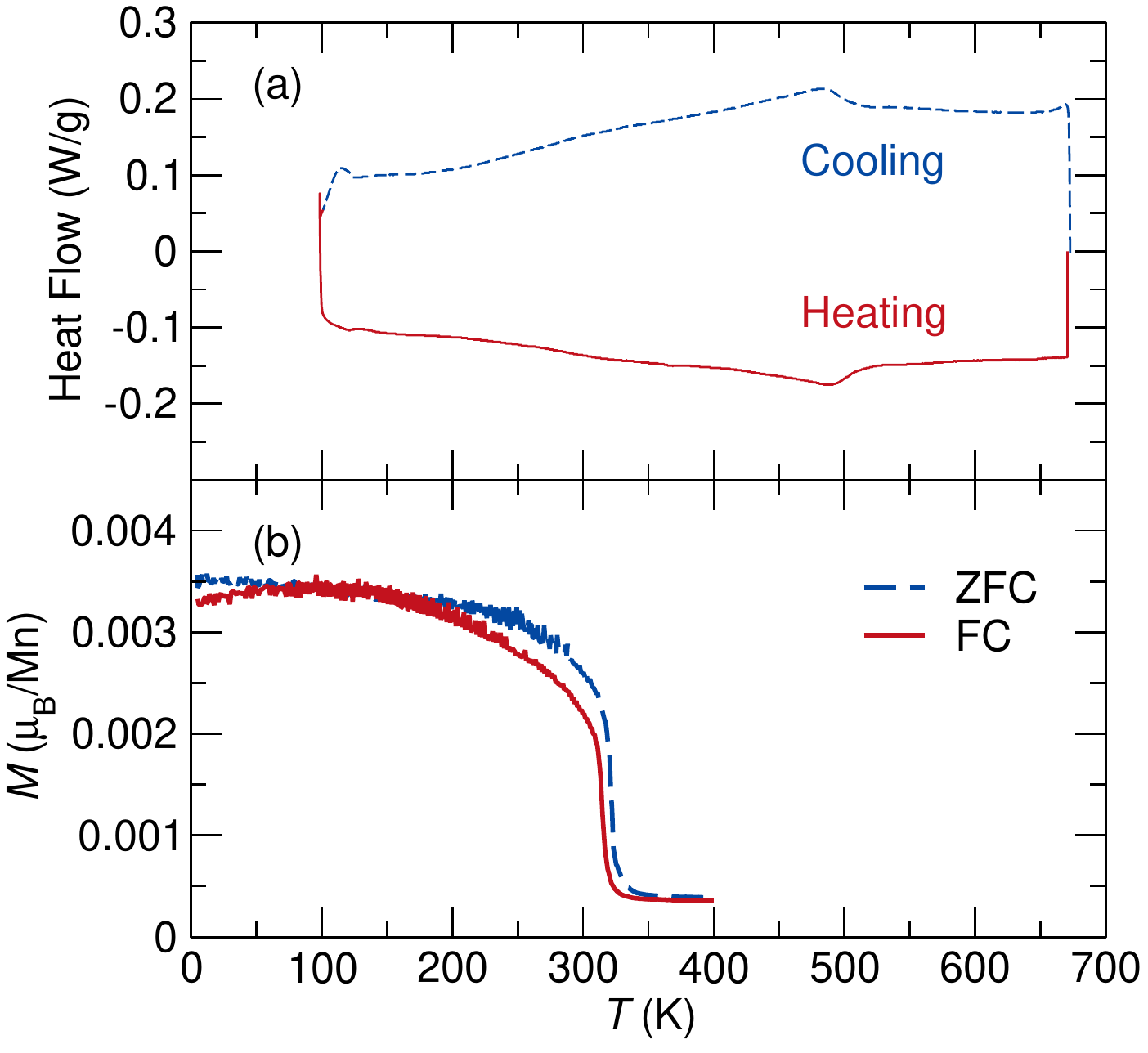} \\
\caption{\label{fig:nearstoich_DSC_SQUID_B}
 Near-stoichiometric \asexcess\ DSC data in (a) and field cooling and zero field cooling magnetization in (b) show two transitions in the near-stoichiometric sample at around 315~K and 490~K. The transition at 490~K corresponds to the T$_N$, while an uncompensated moment develops at 315~K due to a transition to weak ferromagnetism (canting) or subtle phase separation.
} 
\end{figure}


\begin{figure}[h]
\centering\includegraphics[width=0.8\columnwidth]{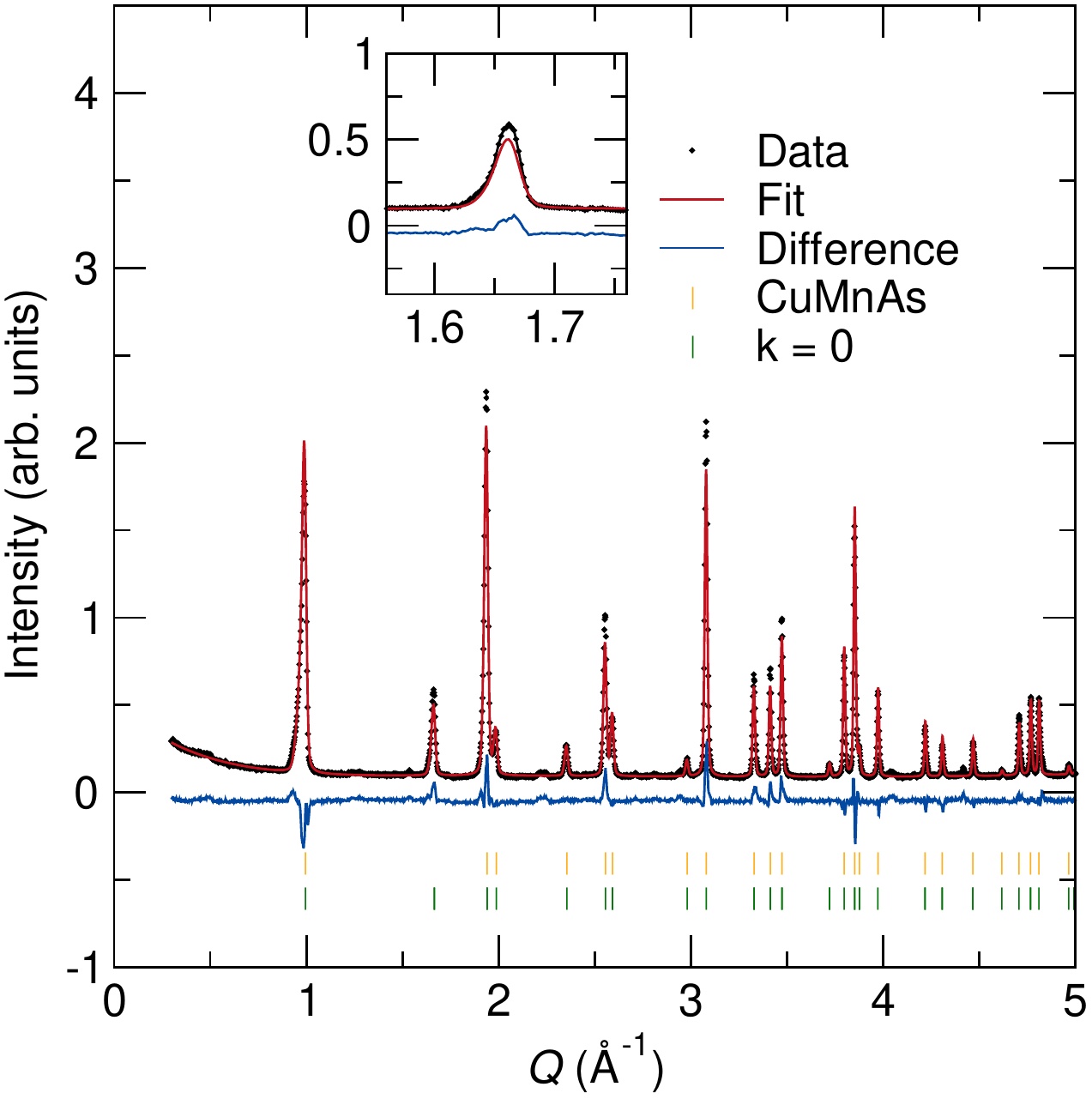} \\
\caption{\label{fig:4K_mag_structure}
Rietveld fit to Echidna NPD data of the near-stoichiometric sample \asexcess\
at 4~K confirms the $Pm'mn$ magnetic space group. The magnetic moment on Mn sites is refined to be 3.73(3) $\mu_B$. 
} 
\end{figure}

\begin{figure}[h]
\centering\includegraphics[width=0.8\columnwidth]{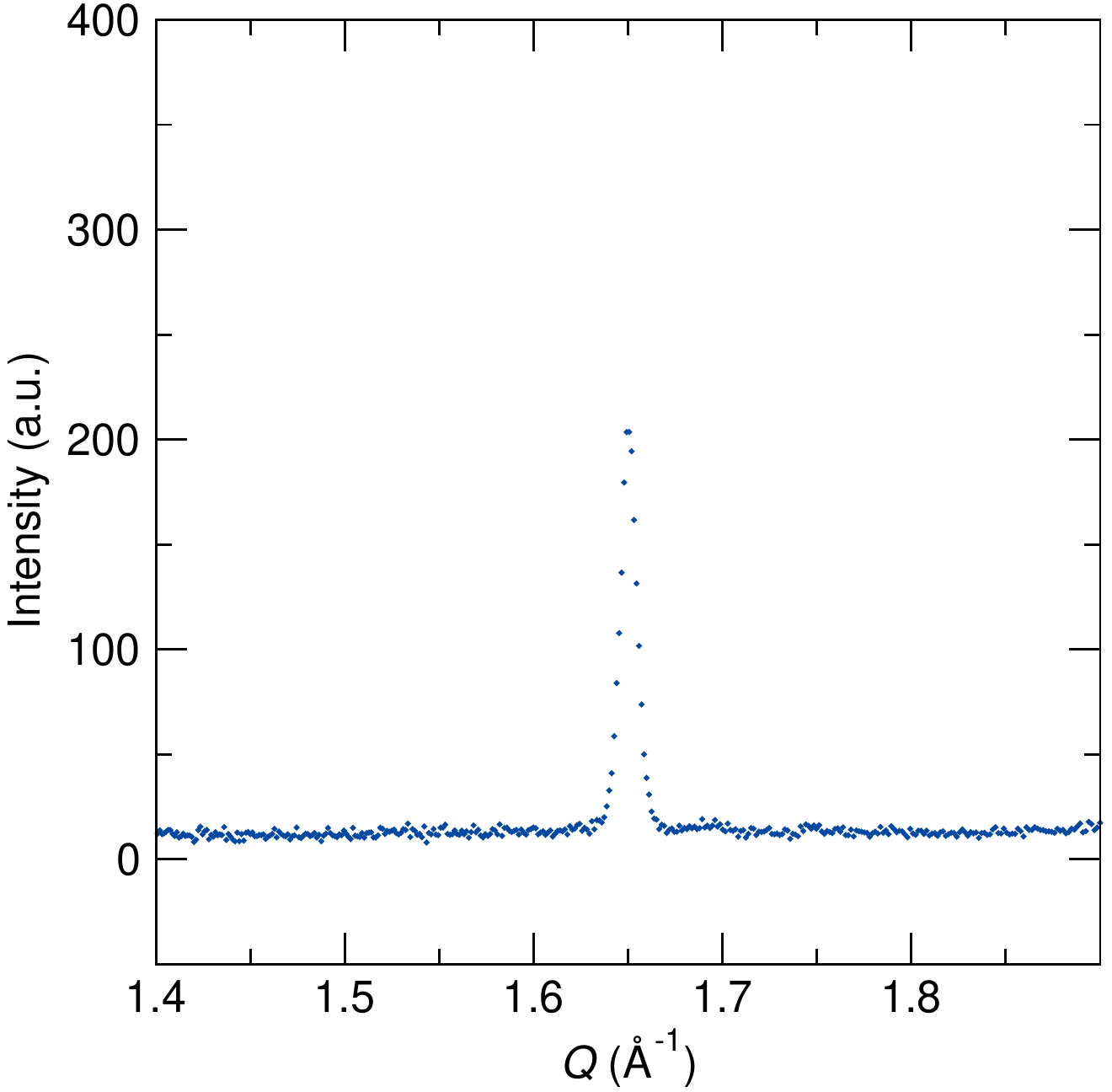} \\
\caption{\label{fig:Cu-rich_POWGEN_300K_100}
300~K POWGEN NPD data showing the presence of the 100 magnetic peak in \cuexcess.
} 
\end{figure}

\begin{table}
\caption{Spacegroups corresponding to the k-maximal subgroups for $P4/nmm1'$ space group and $k=0$ propagation vector.}
\begin{tabular}{|l|l|}
\hline
No. & Group symbol\\
\hline
1 & P4/n'm'm'\\
2 & P4'/n'mm' \\
3 & P4/nm'm' \\
4 & P4'/n'm'm \\
5 & P4'/nmm' \\
6 & P4'/nm'm \\
7 & P4/n'mm \\
8 & P4/nmm \\
9 & Cmm'a' \\
10 & Cm'ma \\
11 & Pmm'n' \\
12 & Pm'mn\\
\hline
\end{tabular}
\end{table}
  
\begin{figure}[h]
\centering\includegraphics[width=0.8\columnwidth]{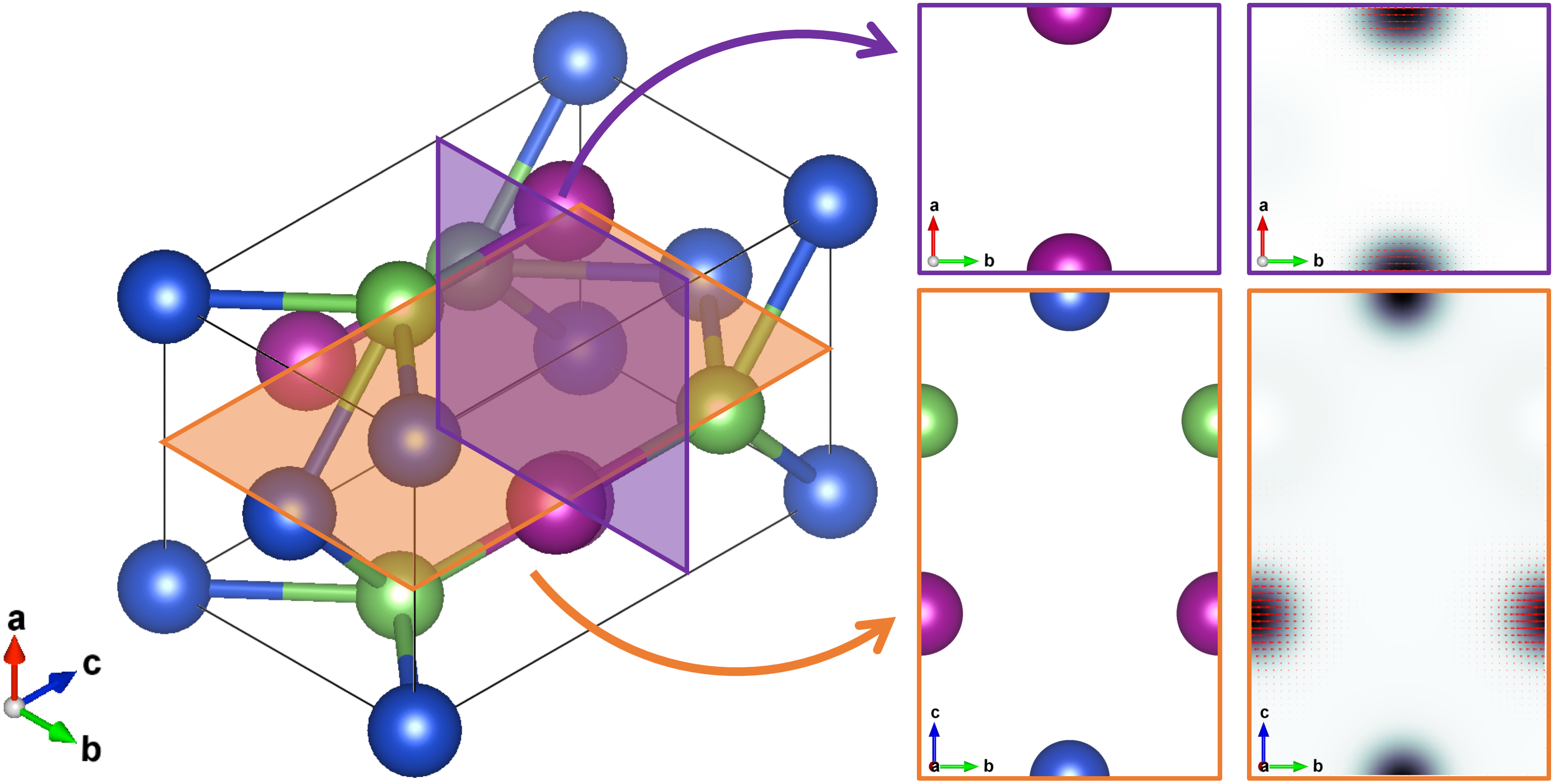} \\
\caption{\label{fig:DFTvis}
The DFT calculated electron charge (Grayscale) and magnetization densities (Red vectors) of tetragonal CuMnAs in $ab$-plane at $z=0.657$ (Purple boxes) and $bc$-plane at $a=0.500$. Color contrast and vector size presents the relative density change. Blue, purple, and green spheres represent the element of Cu, Mn, and As, respectively.
} 
\end{figure}

\begin{figure}[h]
\centering\includegraphics[width=0.96\columnwidth]{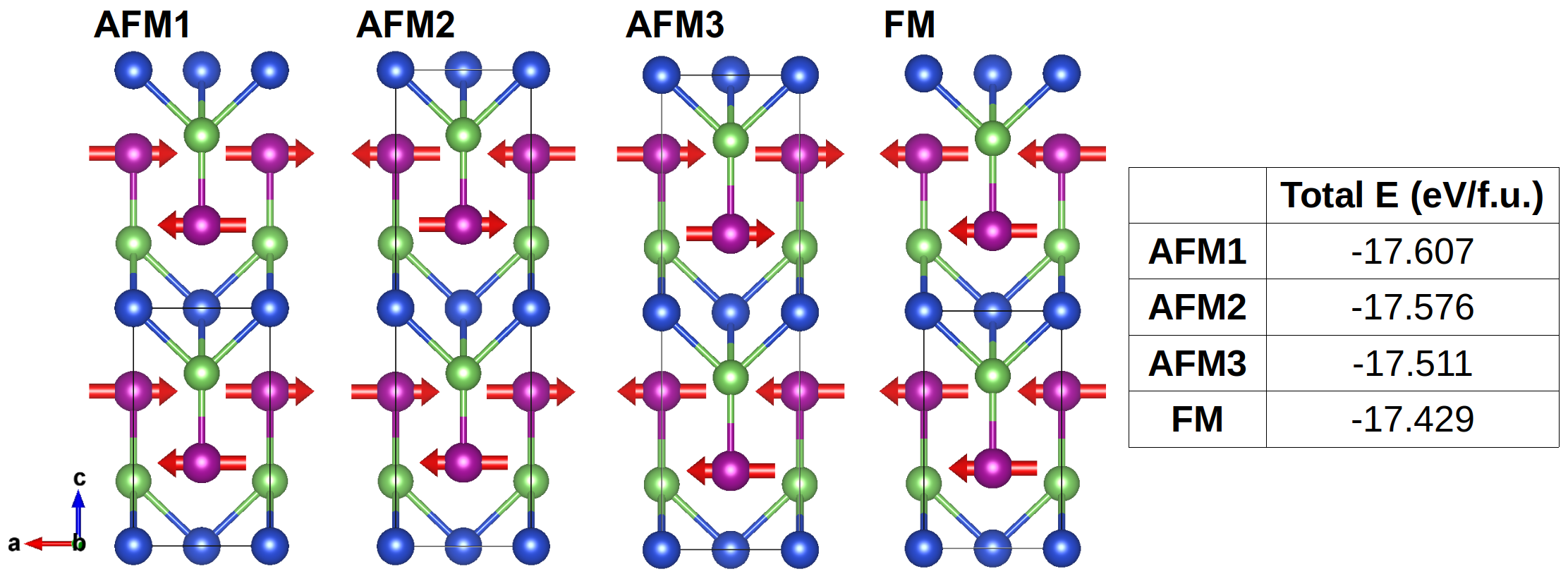} \\
\caption{\label{fig:DFTvis}
Three antiferromagnetic (AFM) configurations and one ferromagnetic (FM) configuration with DFT total energies tabulated for comparison. AFM1 is the ground state structure of CuMnAs which has the lowest total energy per formula unit (f.u.). Solid gray lines show the magnetic unit cells.
} 
\end{figure}


\end{center}

\clearpage

Sequential peak fits shown in Fig. 2(b) and S1 was performed using GSAS-II. The $c$ lattice parameter was extracted by fitting of a pseudo Voigt peak to the 001 reflection in the dataset and results are shown in the tables below. In case of the POWGEN NPD data, the peak position, intensity, Gaussian and Lorentzian peak width parameters for the 001 peak were allowed to refine. For the 11-BM XRD peak fit data in Fig. 2 (b), the Gaussian peak width parameter for the 001 peak was fixed to the instrumental parameters and the other three parameters were allowed to refine. For the Wombat peak fit data in S1, a simple Gaussian peak fitting was done with the Lorentzian peak fit parameters fixed to 0. For all the sequential peak fits, a third order Chebychev polynomial was used to fit the background and the refined values were allowed to copy from one temperature data to the next sequentially. Fig. S9 shows the fit to the 001 peak of the 152~K \cuexcess\ sample measured at POWGEN. Tables S2, S3 and S4 show the refined peak fitting parameters corresponding to the POWGEN NPD, 11-BM XRD and Wombat NPD data of \cuexcess\ respectively.

\begin{center}
\begin{figure}[h]
\centering\includegraphics[width=0.6\columnwidth]{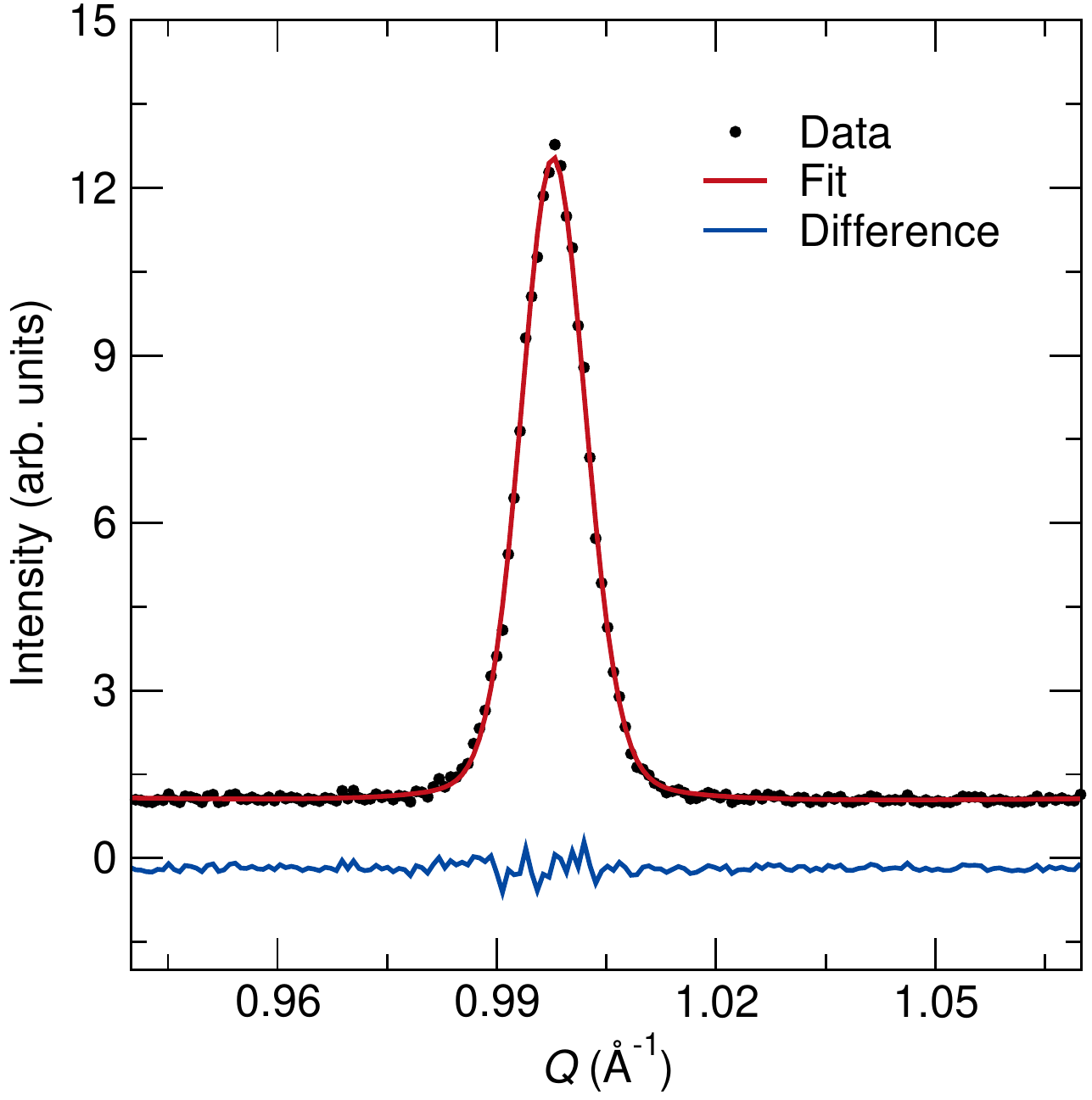} \\
\caption{\label{fig:peakfit}
The peak fit to the 001 peak of the POWGEN NPD data at 152~K.}
\end{figure}
\end{center}

\clearpage

\begin{sidewaystable}
\begin{center}
\caption{Refined sequential peak fitting parameters to the 001 reflection along with their standard deviations for the POWGEN Cu-excess \cumnas\ NPD data shown in Fig. 2(b). }
\begin{tabular}{c|c|c|c|c|c|c|c|c|c}
temp & Rwp & pos0 & esd-pos0 & int0 & esd-int0 & sig0 & esd-sig0 & gam0 & esd-gam0\\
\hline
498.697 & 19.70919995 & 142064.2043 & 11.89517485 & 91.03075408 & 3.172722017 & 261111.8668 & 32730.36639 & 289.631438 & 94.00402322\\
492.118 & 13.54110369 & 142022.9316 & 8.376332891 & 92.26197397 & 2.116154708 & 293682.0572 & 23720.95443 & 278.9033833 & 64.26491834\\
482.03 & 15.09778924 & 142022.1066 & 9.232020068 & 92.01398648 & 2.3077987 & 263327.9765 & 25164.49605 & 340.1299723 & 69.54713756\\
472.059 & 12.5558187 & 142029.2615 & 8.039087335 & 91.68728784 & 1.969934601 & 288853.4697 & 23499.52459 & 356.1407048 & 62.45386417\\
462.058 & 13.69273752 & 142036.5159 & 8.632418293 & 89.81623427 & 2.120571812 & 306329.7096 & 25264.35808 & 277.7943791 & 67.54215898\\
452.059 & 14.52488361 & 142042.0642 & 9.242748057 & 89.11976022 & 2.245279882 & 295945.1682 & 26812.1146 & 307.1378455 & 72.00291804\\
442.058 & 13.14758744 & 142052.7763 & 8.201075086 & 92.25994085 & 2.053191097 & 278420.1242 & 22979.57566 & 320.5502855 & 62.66029647\\
432.058 & 13.39151292 & 142089.467 & 8.593366857 & 100.2217677 & 2.197344135 & 266209.2117 & 23967.23283 & 423.0849619 & 63.81899171\\
422.058 & 13.63456674 & 142121.7202 & 8.64524663 & 102.4950383 & 2.309466379 & 288728.6549 & 24664.71592 & 348.9224182 & 65.36248483\\
412.063 & 12.08780665 & 142135.8681 & 7.570366948 & 107.0243996 & 2.152104344 & 311238.6256 & 22055.32004 & 295.6162067 & 57.91037431\\
402.108 & 12.56172878 & 142163.1174 & 7.953752978 & 115.1970066 & 2.2512583 & 276751.3963 & 21929.95057 & 402.6866303 & 57.03588341\\
392.056 & 11.80367431 & 142182.2056 & 7.406429919 & 115.0788471 & 2.191776553 & 311919.9111 & 21252.28163 & 308.1759318 & 55.13847276\\
382.077 & 11.9935159 & 142183.1995 & 7.362049041 & 120.4085156 & 2.262025715 & 302706.2256 & 20646.3983 & 310.8026878 & 53.78951368\\
371.751 & 10.41975736 & 142205.2772 & 6.392465754 & 120.9675868 & 1.891036256 & 280899.1454 & 17141.77449 & 364.0328098 & 44.63232954\\
363.362 & 12.91860046 & 142194.8479 & 7.853146669 & 126.1723885 & 2.442981183 & 260486.7334 & 20534.89779 & 393.8394959 & 54.26538639\\
350.794 & 11.83335286 & 142215.4703 & 7.412986512 & 131.8806781 & 2.423036751 & 308390.7117 & 20865.67004 & 335.7074678 & 53.23594496\\
341.971 & 10.22148203 & 142222.2726 & 6.269498476 & 134.1329972 & 2.10031602 & 312882.699 & 17599.88764 & 304.1651145 & 45.05911\\
332.061 & 10.52695386 & 142224.9366 & 6.499426145 & 136.3004857 & 2.171009655 & 303511.2201 & 18194.88143 & 352.6541033 & 46.1689538\\
322.066 & 10.92999646 & 142211.0947 & 6.614516374 & 139.224979 & 2.204126961 & 280621.325 & 17356.75226 & 356.6054537 & 44.7361311\\
312.074 & 10.36017591 & 142198.564 & 6.37019055 & 140.9445976 & 2.154896862 & 286832.7771 & 17229.70134 & 380.6751165 & 44.00912864\\
302.082 & 9.596990264 & 142195.3467 & 5.785224313 & 143.2223686 & 2.071439294 & 296868.5662 & 15770.61611 & 322.8685853 & 41.04099938\\
292.069 & 9.521755809 & 142180.7335 & 5.709152175 & 144.1244066 & 2.04792454 & 311356.9425 & 15656.4054 & 280.2254865 & 40.19971928\\
282.063 & 10.37817594 & 142174.5027 & 6.227032556 & 150.9172264 & 2.241212811 & 267834.4901 & 16321.75197 & 390.0768859 & 42.15654282\\
272.056 & 9.273976232 & 142157.6046 & 5.564263522 & 152.5458677 & 2.060504925 & 317670.4759 & 15069.96668 & 291.6131932 & 38.16997117\\
261.993 & 8.436090716 & 142139.5171 & 5.07537608 & 153.3416852 & 1.840642791 & 283253.4929 & 13273.47396 & 357.7666187 & 33.99971475\\
251.985 & 9.11517559 & 142123.862 & 5.454546747 & 154.7994422 & 2.012814595 & 292723.8553 & 14330.05008 & 330.4267415 & 36.69104893\\
242 & 7.817654705 & 142102.9272 & 4.805040842 & 157.5478564 & 1.841567223 & 323867.5524 & 13503.86984 & 308.2787958 & 33.79148516\\
232.01 & 8.347922128 & 142094.1284 & 5.043567996 & 162.8928885 & 1.908216283 & 271400.0774 & 13045.40672 & 396.8823716 & 33.24305185\\
222.015 & 7.819109279 & 142073.7458 & 4.75904787 & 160.7743943 & 1.851393515 & 327412.2318 & 13250.91091 & 290.4404306 & 33.15431454\\
212.028 & 8.17778606 & 142066.3556 & 4.837631829 & 159.0741978 & 1.849344885 & 301349.4494 & 12613.1721 & 293.627302 & 32.34942052\\
202.024 & 8.228743731 & 142057.2846 & 5.005763881 & 165.8895115 & 1.993152128 & 306187.1818 & 13640.77823 & 339.802775 & 34.46648655\\
192.026 & 8.415283195 & 142032.9036 & 5.066331911 & 167.6992859 & 1.97476538 & 298696.9035 & 13333.91623 & 343.5551142 & 33.51005934\\
182.047 & 7.463744237 & 142024.8804 & 4.460787198 & 164.9643845 & 1.766085965 & 316943.9959 & 11990.59891 & 284.6743197 & 30.26655757\\
172.058 & 7.800704689 & 142013.178 & 4.617229936 & 167.4163055 & 1.848451317 & 307473.1461 & 12162.58332 & 292.5265345 & 30.88205289\\
162.08 & 8.482397174 & 141993.9853 & 5.096087829 & 173.8399244 & 2.087653031 & 289897.1563 & 13490.2768 & 362.0933478 & 34.18123743\\
152.068 & 7.823312938 & 141986.611 & 4.691766772 & 174.3034442 & 1.920973621 & 311738.9485 & 12527.16636 & 311.8343925 & 31.35023139\\
143.277 & 9.620936814 & 141967.474 & 5.807960476 & 178.0064684 & 2.44214405 & 306679.0547 & 15577.52104 & 340.1741943 & 39.07347679\\
    \end{tabular}
\end{center}
\end{sidewaystable}

\clearpage

\begin{center}
\setlength\LTleft{-2.52cm}

\begin{longtable}{c|c|c|c|c|c|c|c}
\caption{Refined sequential peak fitting parameters to the 001 reflection along with their standard deviations for the 11-BM Cu-excess \cumnas\ XRD data shown in Fig. 2(b). }\\
temp & Rwp & pos0 & esd-pos0 & int0 & esd-int0 & gam0 & esd-gam0\\
\hline
100 & 15.40898477 & 3.764751113 & 8.35E-05 & 33397.07756 & 382.2453044 & 1.056735425 & 0.0212632189557216\\
105.2 & 17.51546945 & 3.764491436 & 0.000104633 & 34843.33894 & 452.9700072 & 1.22996291 & 0.0273718893032842\\
110.2 & 16.64001045 & 3.764090143 & 9.31E-05 & 36398.23524 & 438.1340359 & 1.152604608 & 0.0242939727914342\\
115.2 & 16.08480568 & 3.763800051 & 8.10E-05 & 40066.97151 & 449.0246268 & 1.028639678 & 0.0204436614313128\\
120.2 & 16.51867079 & 3.763768404 & 8.33E-05 & 43355.88086 & 491.7798408 & 1.057780746 & 0.0210815150360724\\
125.19 & 15.59653616 & 3.763375851 & 7.96E-05 & 43692.66835 & 466.142366 & 1.08234984 & 0.0201860482607935\\
130.19 & 16.28443385 & 3.763254547 & 8.20E-05 & 42699.49297 & 475.9167351 & 1.050960312 & 0.020656638909363\\
135.19 & 15.99554012 & 3.762955582 & 8.21E-05 & 39542.65787 & 444.2611975 & 1.051928087 & 0.020853206888371\\
140.2 & 15.65893372 & 3.76270489 & 7.93E-05 & 40360.71056 & 437.8302686 & 1.050677538 & 0.0200367693083477\\
145.27 & 15.62411654 & 3.76268682 & 7.99E-05 & 38887.88471 & 425.2170108 & 1.045245065 & 0.0201324794950738\\
150.15 & 15.81089208 & 3.762378479 & 8.27E-05 & 36948.46815 & 415.6468142 & 1.06638088 & 0.0210261887152205\\
155.17 & 15.62338487 & 3.762268703 & 8.01E-05 & 36358.49056 & 405.1793984 & 1.029045531 & 0.0204203609270273\\
160.21 & 15.45819148 & 3.762137295 & 7.82E-05 & 36624.75282 & 401.3950352 & 1.01860072 & 0.019818137325659\\
165.2 & 15.81840785 & 3.76182039 & 8.20E-05 & 35941.4254 & 406.7087208 & 1.052284262 & 0.0209886756959273\\
170.19 & 15.29252082 & 3.761747786 & 7.91E-05 & 36207.27257 & 395.007062 & 1.043800044 & 0.0201340126245682\\
175.2 & 15.24794647 & 3.761338738 & 7.73E-05 & 35582.816 & 386.2799501 & 1.015979053 & 0.019655253890676\\
180.2 & 16.15203071 & 3.761274717 & 8.07E-05 & 35760.24772 & 410.4768422 & 0.995531207 & 0.020433692799927\\
185.21 & 15.36129759 & 3.76115447 & 7.82E-05 & 35568.75545 & 391.3019804 & 1.020036428 & 0.0198986975224395\\
190.19 & 15.13780374 & 3.761107791 & 7.76E-05 & 36015.07521 & 387.9481262 & 1.035911341 & 0.0197357621932927\\
195.19 & 15.08483973 & 3.760780175 & 7.64E-05 & 35995.75982 & 386.4955387 & 1.008597481 & 0.0191879155602154\\
200.2 & 14.96575267 & 3.760576062 & 7.63E-05 & 36070.41271 & 384.1977905 & 1.02802052 & 0.0193391758595985\\
205.19 & 15.15841732 & 3.760436746 & 7.74E-05 & 35999.53596 & 387.4603528 & 1.030211139 & 0.0196103502064985\\
210.18 & 14.9717003 & 3.760286392 & 7.63E-05 & 34973.29904 & 375.8489549 & 1.011371857 & 0.0193170880132166\\
215.13 & 14.91454304 & 3.760012086 & 7.59E-05 & 35195.77941 & 377.0560308 & 1.0075744 & 0.019098232808584\\
220.17 & 15.89122801 & 3.760077826 & 8.03E-05 & 35285.51522 & 399.7829089 & 1.006310965 & 0.0203254030717933\\
225.16 & 15.27657253 & 3.759864671 & 7.72E-05 & 35932.01875 & 389.6442907 & 1.012736164 & 0.0194062512326813\\
230.15 & 14.90683396 & 3.759650396 & 7.54E-05 & 35793.32411 & 379.2563944 & 1.014432984 & 0.0190559684970362\\
235.16 & 15.28680237 & 3.759591037 & 7.86E-05 & 35097.45213 & 385.5297731 & 1.017278298 & 0.0197364155761368\\
240.14 & 15.06844322 & 3.759502947 & 7.75E-05 & 35512.11483 & 379.6821172 & 1.041441279 & 0.0195960612897471\\
245.13 & 15.57641979 & 3.759513688 & 7.91E-05 & 34863.22697 & 388.8833498 & 0.99873588 & 0.0198386838705586\\
250.12 & 15.35333721 & 3.759542482 & 7.90E-05 & 35367.9909 & 387.9153201 & 1.028026098 & 0.0198406781274231\\
255.09 & 15.72595555 & 3.759588805 & 7.88E-05 & 35225.98073 & 392.3778012 & 0.991575466 & 0.019635686402649\\
260.1 & 15.55673461 & 3.759555153 & 7.86E-05 & 35277.69165 & 390.6356289 & 1.001028798 & 0.0195966164539501\\
265.08 & 15.63534499 & 3.759596244 & 7.85E-05 & 35062.66853 & 391.9005865 & 0.992035516 & 0.0197869791847228\\
270.09 & 15.95765712 & 3.759389735 & 8.00E-05 & 35234.57073 & 401.9811962 & 0.990931752 & 0.0200992258818481\\
275.11 & 15.42941352 & 3.759290621 & 7.65E-05 & 35201.4097 & 386.9702441 & 0.974941054 & 0.0191008186566161\\
280.1 & 15.16564964 & 3.759033566 & 7.61E-05 & 35656.53038 & 383.1062221 & 1.001045446 & 0.0189728894043922\\
285.11 & 15.67590988 & 3.758862529 & 7.93E-05 & 35225.51762 & 394.0171707 & 1.007101466 & 0.019969634736104\\
290.12 & 15.25928936 & 3.758650651 & 7.69E-05 & 34518.50687 & 378.6811582 & 0.984482722 & 0.0191758043680333\\
295.09 & 14.84767236 & 3.758528724 & 7.53E-05 & 35119.78487 & 371.6292695 & 1.008052226 & 0.0187802482874206\\
300.15 & 15.24209309 & 3.758358833 & 7.57E-05 & 34298.3956 & 374.8426533 & 0.963329949 & 0.0187970868737863\\
305.09 & 14.79775567 & 3.758172646 & 7.24E-05 & 35226.25348 & 368.8454206 & 0.960666297 & 0.0178306333997266\\
310.12 & 14.3282203 & 3.758110765 & 7.05E-05 & 35756.39404 & 361.7063091 & 0.976204573 & 0.0175457230138955\\
315.83 & 13.85510229 & 3.757953688 & 6.70E-05 & 36180.04514 & 350.8458103 & 0.955159639 & 0.0164512960163215\\
320.02 & 11.37940007 & 3.758176236 & 5.16E-05 & 37681.50513 & 291.4604792 & 0.892724354 & 0.0123091897394304\\
325.2 & 11.17482615 & 3.758132607 & 4.71E-05 & 38691.74526 & 287.4427031 & 0.824236423 & 0.0111072794521553\\
329.9 & 10.33446143 & 3.758297639 & 4.16E-05 & 38868.44885 & 263.8618366 & 0.768996066 & 0.00954610950563794\\
334.96 & 9.956208556 & 3.758436172 & 3.89E-05 & 39164.77307 & 254.2327092 & 0.734596084 & 0.00877774699703509\\
340.25 & 9.403174501 & 3.75858751 & 3.53E-05 & 40110.94539 & 242.5647047 & 0.702700273 & 0.00789560242962205\\
345.27 & 9.76542834 & 3.758552612 & 3.54E-05 & 40128.74931 & 251.1607097 & 0.657928367 & 0.00777902094197381\\
350.23 & 9.840460907 & 3.758665684 & 3.51E-05 & 40266.36631 & 252.7593054 & 0.6441682 & 0.00769741474775996\\
355.38 & 10.3248673 & 3.758876461 & 3.69E-05 & 40047.20736 & 264.8910918 & 0.641257556 & 0.00803429397476061\\
360.33 & 10.1947289 & 3.759016254 & 3.56E-05 & 40124.10888 & 261.6940894 & 0.617763071 & 0.00773605736756989\\
365.26 & 10.51603188 & 3.759369846 & 3.63E-05 & 43080.01186 & 283.7433985 & 0.624634581 & 0.00784401144994696\\
370.01 & 10.79352108 & 3.759564117 & 3.71E-05 & 43019.45781 & 291.1149261 & 0.616661036 & 0.00797155134049598\\
375.1 & 10.27845062 & 3.759835284 & 3.55E-05 & 43039.78161 & 277.0625168 & 0.623949946 & 0.00766775822245959\\
380.12 & 11.12539058 & 3.760080812 & 3.76E-05 & 42642.05213 & 296.98925 & 0.596441441 & 0.00802700976218355\\
385.3 & 10.54782969 & 3.760511148 & 3.54E-05 & 44358.55929 & 289.8647602 & 0.598951931 & 0.00754289577463892\\
389.97 & 10.7942556 & 3.760867239 & 3.57E-05 & 45656.56665 & 302.4390332 & 0.591327851 & 0.00753935412423171\\
394.96 & 10.65902673 & 3.761224217 & 3.57E-05 & 46444.08129 & 303.4439861 & 0.605295357 & 0.00757227221748354\\
399.96 & 9.695928737 & 3.761745525 & 3.25E-05 & 46470.25039 & 275.6935058 & 0.612042346 & 0.00696908292538409\\
405 & 10.84712877 & 3.761917685 & 3.73E-05 & 46589.59889 & 310.5796502 & 0.630435007 & 0.0079598784973985\\
410.13 & 11.44325611 & 3.762422128 & 4.03E-05 & 46185.21545 & 326.1164035 & 0.652977742 & 0.00864539826926886\\
414.99 & 11.63158536 & 3.762842638 & 4.16E-05 & 45638.00518 & 328.6148828 & 0.665555947 & 0.00890745305045242\\
420.15 & 11.16260733 & 3.763367875 & 4.03E-05 & 45392.87451 & 314.2806841 & 0.673528959 & 0.00864243855463501\\
425.14 & 11.33650796 & 3.763681376 & 4.01E-05 & 45231.48778 & 317.557912 & 0.654379643 & 0.00861016789831714\\
430.17 & 10.64769826 & 3.764006894 & 3.65E-05 & 44311.49973 & 292.9864259 & 0.619244665 & 0.00783234922720501\\
435.25 & 10.4220952 & 3.764322265 & 3.43E-05 & 44086.44815 & 284.964483 & 0.580715505 & 0.00737227895145399\\
440.14 & 10.10917038 & 3.764435604 & 3.21E-05 & 43825.11119 & 273.1317943 & 0.552451109 & 0.00685338475593841\\
445.26 & 9.462606348 & 3.764537493 & 2.91E-05 & 43876.79674 & 255.894621 & 0.523602011 & 0.00621245579953472\\
450.32 & 9.054984747 & 3.76462077 & 2.72E-05 & 43587.34706 & 243.3447585 & 0.501106865 & 0.00577838893756978\\
455.37 & 10.19865966 & 3.76465697 & 3.00E-05 & 43584.72908 & 272.8056885 & 0.483380883 & 0.00632642902762149\\
460.46 & 9.539136882 & 3.764881769 & 2.74E-05 & 43348.93234 & 254.0139352 & 0.462112587 & 0.00580362841825301\\
465.13 & 8.538396941 & 3.764687691 & 2.40E-05 & 43008.46016 & 224.9367321 & 0.444243371 & 0.00504165114184438\\
470.33 & 8.95335558 & 3.764662579 & 2.51E-05 & 42978.81798 & 235.9361856 & 0.443117583 & 0.00529368839531341\\
475.34 & 8.864990657 & 3.764586369 & 2.48E-05 & 42525.91916 & 231.4459592 & 0.441554305 & 0.00523668190338918\\
480.23 & 9.684587259 & 3.764558131 & 2.75E-05 & 41553.38031 & 249.6543345 & 0.445155289 & 0.00581408325888385\\
    \end{longtable}
\end{center}

\clearpage

\begin{center}
\setlength\LTleft{-2.35cm}

\begin{longtable}{c|c|c|c|c|c|c|c}
\caption{Refined sequential peak fitting parameters to the 001 reflection along with their standard deviations for the Wombat \cumnas\ XRD data shown in Fig. S1. }\\
temp & Rwp & pos0 & esd-pos0 & int0 & esd-int0 & sig0 & esd-sig0\\
\hline
15.0093 & 2.854853291 & 21.92272617 & 0.004092967 & 380177.2816 & 3876.632204 & 1409.892651 & 37.9223619662951\\
20.0055 & 2.889402599 & 21.92123854 & 0.004129074 & 381807.8208 & 3929.75035 & 1405.989165 & 38.1920813416562\\
25.0033 & 2.850599506 & 21.91809677 & 0.004037098 & 381601.0628 & 3847.207311 & 1372.241621 & 36.7900558642892\\
29.9996 & 3.194539358 & 21.91540408 & 0.004524125 & 382981.1402 & 4326.544183 & 1380.85982 & 41.4444304570222\\
34.9858 & 3.1083564 & 21.91464761 & 0.004369889 & 385468.9417 & 4203.044569 & 1367.812042 & 39.7623053238669\\
39.972 & 3.020827136 & 21.91679919 & 0.004235903 & 384742.7602 & 4077.768833 & 1351.633511 & 38.3198098787686\\
45.0021 & 2.962663943 & 21.91407696 & 0.004158933 & 382627.4128 & 3987.397808 & 1334.758365 & 37.2713868867435\\
50.1008 & 2.918840446 & 21.91435206 & 0.004105848 & 384063.2479 & 3945.416869 & 1346.718126 & 37.0960468741626\\
55.1986 & 3.05890192 & 21.91660673 & 0.004283034 & 385362.1142 & 4132.191503 & 1347.319177 & 38.7453434406783\\
60.3007 & 3.200020699 & 21.91492761 & 0.004484061 & 384148.0091 & 4312.150144 & 1350.808372 & 40.5409139636033\\
65.353 & 3.104185628 & 21.91540753 & 0.00436497 & 382671.3568 & 4179.189 & 1353.0041 & 39.5071947240561\\
70.3992 & 3.047770618 & 21.91329811 & 0.0042774 & 383771.4632 & 4102.610743 & 1355.768487 & 38.71621404171\\
75.4216 & 3.240342504 & 21.91210335 & 0.004543773 & 381964.8237 & 4347.690581 & 1342.040941 & 40.9331065639604\\
80.4594 & 3.029876405 & 21.91383898 & 0.00424803 & 381918.1549 & 4068.220181 & 1330.987252 & 38.1283675839476\\
85.4615 & 2.906157849 & 21.91103928 & 0.004103658 & 382756.1692 & 3921.372877 & 1365.127178 & 37.3697206375552\\
90.4671 & 2.960410939 & 21.91211798 & 0.004159066 & 382229.6195 & 3973.106021 & 1353.047069 & 37.5725594541275\\
95.4614 & 3.084925431 & 21.91201565 & 0.004333198 & 379858.0393 & 4126.284513 & 1336.187812 & 38.9965119258992\\
100.464 & 3.224865575 & 21.91060705 & 0.004538936 & 379571.9389 & 4319.643771 & 1337.93226 & 40.8362187651514\\
105.466 & 3.154655005 & 21.9085769 & 0.004416421 & 381270.7463 & 4215.504609 & 1339.337345 & 39.7134783122608\\
110.458 & 3.123204383 & 21.91015931 & 0.00438404 & 380006.5094 & 4177.221012 & 1333.233224 & 39.3947902233174\\
115.46 & 3.086357193 & 21.90904737 & 0.004287301 & 381487.0689 & 4111.556366 & 1306.201356 & 38.0018443443138\\
120.453 & 3.03169838 & 21.90847347 & 0.004254293 & 380606.0891 & 4060.333125 & 1331.33731 & 38.1069568155862\\
125.45 & 2.904302228 & 21.9082303 & 0.004077323 & 381516.8913 & 3903.344892 & 1328.62168 & 36.5127003250654\\
130.395 & 3.113396674 & 21.90653373 & 0.004397682 & 379739.2838 & 4183.643499 & 1341.631819 & 39.6955642777154\\
135.4 & 2.979391498 & 21.90657787 & 0.004218195 & 379314.4581 & 4013.664378 & 1338.385134 & 38.0252225833893\\
140.407 & 3.011572733 & 21.90496695 & 0.004260674 & 379305.9327 & 4053.408699 & 1333.934024 & 38.1862056077721\\
145.398 & 2.903480603 & 21.90254707 & 0.004098893 & 378639.6734 & 3893.61042 & 1328.294297 & 36.67425980905\\
150.391 & 3.141144859 & 21.90347539 & 0.004440743 & 378642.7398 & 4221.752343 & 1331.676068 & 39.9608650356348\\
155.392 & 2.947179396 & 21.90243155 & 0.004199122 & 376880.9393 & 3964.852598 & 1353.048038 & 38.1170827821062\\
160.385 & 2.741308279 & 21.90353151 & 0.003910623 & 374622.8352 & 3670.242136 & 1351.985801 & 35.3736275985411\\
165.388 & 3.045980775 & 21.89915048 & 0.004347885 & 374560.0467 & 4080.241446 & 1354.798398 & 39.4989194639846\\
170.382 & 3.004675556 & 21.90028009 & 0.004290812 & 373055.1217 & 4013.773015 & 1344.883798 & 38.8708868389681\\
175.384 & 3.099432953 & 21.89627003 & 0.004411228 & 373647.584 & 4138.081573 & 1341.17985 & 39.8573728820509\\
180.379 & 3.025372645 & 21.89597268 & 0.004324805 & 369870.3355 & 4021.921602 & 1330.271517 & 38.8691079541265\\
185.379 & 3.022287339 & 21.89537552 & 0.004310161 & 370571.2935 & 4018.425894 & 1326.509137 & 38.5856969323809\\
190.371 & 2.957882457 & 21.89319274 & 0.004257737 & 366598.8348 & 3924.826009 & 1336.077939 & 38.3130631936896\\
195.371 & 2.666134883 & 21.89339164 & 0.003826681 & 367572.9857 & 3536.584169 & 1331.551327 & 34.3568228131088\\
200.363 & 3.022331187 & 21.8917403 & 0.004344816 & 367046.5398 & 4008.148832 & 1340.053967 & 39.2581934774517\\
205.365 & 2.942678471 & 21.88945652 & 0.004228598 & 365653.0863 & 3889.088912 & 1328.763045 & 37.8456451765898\\
210.359 & 3.142429282 & 21.88730028 & 0.004511471 & 362669.5099 & 4128.455997 & 1312.839229 & 40.2281212042576\\
215.356 & 2.759341036 & 21.88779157 & 0.003992372 & 362922.1378 & 3642.037114 & 1334.617528 & 35.8856186500462\\
220.347 & 2.910806245 & 21.88348019 & 0.004222324 & 359455.6963 & 3822.910119 & 1328.112706 & 37.8814569908907\\
225.338 & 3.085798411 & 21.88207963 & 0.004481339 & 358701.1932 & 4048.054899 & 1334.852229 & 40.3722782204839\\
230.337 & 2.88395726 & 21.88295219 & 0.004216967 & 357947.3247 & 3795.19877 & 1343.809583 & 38.1173580792114\\
235.33 & 2.929767064 & 21.88156014 & 0.004258298 & 355854.7966 & 3825.723507 & 1318.417643 & 38.0861008092353\\
240.329 & 2.92741548 & 21.87885815 & 0.004275616 & 353332.4292 & 3814.423094 & 1320.265108 & 38.3253702059497\\
245.321 & 2.899819226 & 21.87562074 & 0.004240178 & 351243.4548 & 3763.899416 & 1312.270088 & 37.8440356334742\\
250.323 & 2.862217096 & 21.87589249 & 0.004203781 & 350170.8836 & 3719.287212 & 1321.09425 & 37.6203412904972\\
255.316 & 2.745286936 & 21.87731852 & 0.004075167 & 348099.3341 & 3574.846609 & 1347.133903 & 36.9380093944504\\
260.316 & 2.939230927 & 21.87400092 & 0.004345696 & 346630.8144 & 3804.11437 & 1326.941593 & 39.001908826738\\
265.312 & 2.754428728 & 21.87259949 & 0.004059079 & 345371.888 & 3543.638049 & 1316.721865 & 36.2840881912612\\
270.311 & 2.754933586 & 21.87319899 & 0.004092339 & 342521.4845 & 3542.856953 & 1319.886353 & 36.6347300042819\\
275.302 & 2.724900267 & 21.87006678 & 0.004054851 & 338628.1274 & 3480.411953 & 1308.378449 & 36.1480581901135\\
280.303 & 2.680131804 & 21.86707639 & 0.003997963 & 338329.6423 & 3429.523831 & 1309.846211 & 35.6528476792393\\
285.293 & 2.803512333 & 21.86638335 & 0.004198631 & 335154.4011 & 3573.197126 & 1307.710865 & 37.4860682994106\\
290.292 & 2.479133314 & 21.86410866 & 0.003734317 & 332770.9736 & 3155.090642 & 1304.847259 & 33.2061367396018\\
295.283 & 2.734869094 & 21.86076684 & 0.004130657 & 333300.5115 & 3488.431956 & 1325.469799 & 37.1813905946469\\
300.283 & 2.731745932 & 21.86018328 & 0.004134237 & 331006.5159 & 3470.276173 & 1321.77425 & 37.1392923400313\\

    \end{longtable}
\end{center}

\clearpage

\begin{table}
\caption{Rietveld refinement parameters for the room temperature 11-BM XRD data of \cuexcess\ sample shown in Fig. 1(a). The figures of merit are wR =   20.34\%, chi$^2$ = 119544, GOF = 2.19.}
\begin{tabular}{c|c c c c c c c c}
\hline
names & Scale & DisplaceX & DisplaceY\\
\hline
values & 6.6110 & -79.4442 & 1029.7001\\
sig & 0.0224 & 2.9959 & 28.2396\\
\hline
\\
\hline
names & bg1 & bg2 & bg3 & bg4 & bg5 & bg6 & bg7 & bg8\\
\hline
values & 40.06 & -59.41 & -34.54 & 921.1 & 533.6 & -4216 & -1416 & 8130\\
sig & 0.2233 & 1.401 & 3.767 & 13.55 & 14.18 & 35.74 & 10.72 & 16.14\\
\hline
\\
\hline
names & bg9 & bg10 & bg11 & bg12\\
\hline
values & 1229 & -6971 & -305.4 & 2187\\
sig & 15.62 & 40.03 & 16.09 & 32.68\\
\hline
\\
& CuMnAs \\
\hline
names & a & b & c & Volume & Size & Mustrain\\
\hline
values & 3.799452 & 3.799452 & 6.297675 & 90.912 & 0.66317 & 2884.1\\
sig & 0.000051 & 0.000051 & 0.000083 & 0.003 & 0.0355 & 32.2\\
\hline
\\
\hline
name & x & y & z & frac & Uiso \\
\hline
As1\\
values & 0.25000 & 0.25000 & 0.23702 & 1.000 & 0.00583\\
sig & & & 0.00023 & & 0.00041\\
Cu1\\
values & -0.25000 & 0.25000 & 0.50000 & 1.000 & 0.00864\\
sig & & & & & 0.00033\\
Mn1\\
values & 0.25 & 0.25 & -0.16672 & 0.820 & 0.01901\\
sig & & & 0.00029 & 0.00 & 0.00072\\
Cu2\\
values & 0.25 & 0.25 & -0.16672 & 0.180 & 0.01901\\
sig & & & 0.00029 & 0.00 & 0.00072\\
\end{tabular}
\end{table}

\clearpage

\setlength\LTleft{-1cm}
\begin{longtable}{c|c c c c c c c c}
\caption{Rietveld refinement parameters for the 500~K \cuexcess\ POWGEN NPD data shown in Fig. 1(b). The figures of merit are wR = 8.83\%, chi$^2$ = 43085.9, GOF = 3.66. The sum of the phase fractions of the two phases was constrained to be 1. The following atoms were constrained to be equivalent - CuMnAs:Mn1 = CuMnAs:Cu2.}\\
\hline
names & Scale & Zero & difA & difB & difC\\
\hline
values & 26472.4699 & 43.497655 & 4.050819 & -19.982011 & 22542.803876\\
sig & 124.0812 & 3.016729 & 0.426505 & 1.301858 & 2.116480\\
\hline
\\
\hline
names & X & Y & Z\\
\hline
values & -16.761112 & 2.851714 & 11.524797\\
sig & 0.891635 & 6.706909 & 1.281101\\
\hline
\\
\hline
names & bg1 & bg2 & bg3 & bg4 & bg5 & bg6\\
\hline
values & 12.12 & -1.67 & -7.61 & 23.96 & 12.67 & -32.92\\
sig & 0.2251 & 0.636 & 1.621 & 2.313 & 1.96 & 2.241\\
\hline
\\
& CuMnAs\\
\hline
names & a & b & c & Volume & Size & Mustrain & Phase fraction\\
\hline
values & 3.831959 & 3.831959 & 6.293129 & 92.408 & 0.98788 & 990.4 & 0.98988\\
sig & 0.000111 & 0.000111 & 0.000173 & 0.008 & 2.8909 & 39.8 & 0.00063\\
\hline
\\
\hline
name & x & y & z & frac & Uiso \\
\hline
As1\\
values & 0.25000 & 0.25000 & 0.23671 & 1.000 & 0.01261\\
sig & & & 0.00027 & & 0.00038\\
Cu1\\
values & -0.25000 & 0.25000 & 0.50000 & 1.000 & 0.01783\\
sig & & & & & 0.00025\\
Mn1\\
values & 0.25000 & 0.25000 & -0.17123 & 0.813 & 0.00916\\
sig & & & 0.00097 & 0.004 & 0.00134\\
Cu2\\
values & 0.25000 & 0.25000 & -0.17123 & 0.187 & 0.00916\\
sig & & & 0.00097 & 0.004 & 0.00134\\
\hline
\\
& MnO\\
\hline
names & a & b & c & Volume & Size & Mustrain & Phase fraction\\
\hline
values & 4.452986 & 4.452986 & 4.452986 & 88.299 & 1.0 & 1159.5 & 0.01012\\
sig & 0.000490 & 0.000490 & 0.000490 & 0.029 & & 417.6 & 0.00063\\
\hline
\\
\hline
name & x & y & z & frac & Uiso \\
\hline
Mn1\\
values & 0.00000 & 0.00000 & 0.00000 & 1.000 & 0.00781\\
sig \\
O1\\
values & 0.50000 & 0.50000 & 0.50000 & 1.000 & 0.00912\\
sig \\
\end{longtable}

\clearpage

\setlength\LTleft{-0.5cm}
\begin{longtable}{c|c c c c c c c c}
\caption{Rietveld refinement parameters for the 500~K \mnexcess\ POWGEN NPD data shown in Fig. 5. The figures of merit are wR = 8.55\%, chi$^2$ = 44425.8, GOF = 3.68. The sum of all the structural phases were constrained to be 1. The following atoms were contrained to be equivalent - CuMnAs-1:Cu1 = CuMnAs-1:Mn2, CuMnAs-2:Mn1 = CuMnAs-2:Cu4 = k=[001/2]:Mn1\_0, CuMnAs-2:Mn2 = CuMnAs-2:Cu5 = k=[001/2]:Mn2\_1. Crystallite size, mustrain and lattice parameters of CuMnAs-2 and k=[001/2] were constrained to be equal.}\\
\hline
names & Scale & Absorption\\
\hline
values & 23483.6778 & 0.0148\\
sig & 359.7159 & 0.0022\\
\hline
\\
\hline
names & bg1 & bg2 & bg3 & bg4 & bg5 & bg6\\
\hline
values & 18.04 & 0.3393 & -12.76 & 14.14 & 13.25 & -19.38\\
sig & 0.2421 & 0.6309 & 1.797 & 2.335 & 2.147 & 2.355\\
\hline
\\
& CuMnAs-1\\
\hline
names & a & b & c & Volume & Size & Mustrain & Phase fraction\\
\hline
values & 3.850603 & 3.850603 & 6.317257 & 93.667 & 2.65410 & 3135.4 & 0.69521\\
sig & 0.000098 & 0.000098 & 0.000213 & 0.004 & 2.9726 & 75.5 & 0.00514\\
\hline
\\
\hline
name & x & y & z & frac & Uiso \\
\hline
As1\\
values & 0.25000 & 0.25000 & 0.24283 & 1.000 & 0.01236\\
sig & & & 0.00059 & & 0.00071\\
Cu1\\
values & -0.25000 & 0.25000 & 0.50000 & 0.795 & 0.01537\\
sig & & & & 0.006 & 0.00096\\
Mn1\\
values & 0.25000 & 0.25000 & -0.16616 & 1.000 & 0.02033\\
sig & & & 0.00099 & & 0.00126\\
Mn2\\
values & -0.25000 & 0.25000 & 0.50000 & 0.205 & 0.01537\\
sig & & & & 0.006 & 0.00096\\
\hline
\\
& CuMnAs-2\\
\hline
names & a & b & c & Volume & Size & Mustrain & Phase fraction\\
\hline
values & 3.80904 & 3.80904 & 6.30661 & 91.501 & 0.38432 & 2623.2 & 0.28625\\
sig & & & & & 0.1298 & 151.3 & 0.00744\\
\hline
\\
\hline
names & D11 & D33\\
\hline
values & 2.076e-06 & -3.62e-06\\
sig & 5.472e-06 & 2.155e-06\\
\hline
\\
\hline
name & x & y & z & frac & Uiso \\
\hline
As1\\
values & 0.25000 & 0.25000 & 0.73340 & 1.000 & 0.01089\\
sig & & & 0.00137 & & 0.00175\\
Mn1\\
values & 0.75000 & 0.25000 & 0.00000 & 0.957 & 0.00291\\
sig & & & & 0.041 & 0.00365\\
Mn2\\
values & 0.25000 & 0.25000 & 0.32696 & 0.954 & 0.00863\\
sig & & & 0.00223 & 0.046 & 0.00453\\
Cu4\\
values & 0.75000 & 0.25000 & 0.00000 & 0.043 & 0.00291\\
sig & & & &  0.041 & 0.00365\\
Cu5\\
values & 0.25000 & 0.25000 & 0.32696 & 0.046 & 0.00863\\
sig & & & 0.00223 & 0.046 & 0.00453\\
\hline
\\
& $k = [0 0 1/2]$\\
\hline
names & a & b & c & Volume & Size & Mustrain & Phase fraction\\
\hline
values & 3.80904 & 3.80904 & 12.61323 & 183.003 & 0.38432 & 2623.2 & 0.14312\\
sig & & & & & 0.1298 & 151.3 & 0.00372 \\
\hline
\\
\hline
names & D11 & D22 & D33\\
\hline
values & 2.076e-06 & 2.076e-06 & -3.62e-06\\
sig & 5.472e-06 & 5.472e-06 & 2.155e-06\\
\hline
\\
\hline
name & x & y & z & frac & Uiso\\
\hline
Mn1\_0\\
values & 0.25000 & 0.75000 & 0.00000 & 0.957 & 0.00291\\
sig & & & & 0.041 & 0.00365\\
Mn2\_1\\
values & 0.75000 & 0.75000 & 0.16348 & 0.954 & 0.00863\\
sig & & & 0.00112 & 0.046 & 0.00453\\
\hline
\\
\hline
name & My \\
\hline
Mn1\_0\\
values & 0.946 \\
sig & 0.094\\
Mn2\_1\\
values & -2.872\\
sig & 0.138\\
\hline
\\
& MnO\\
\hline
names & a & b & c & Volume & Size & Mustrain & Phase fraction\\
\hline
values & 4.453855 & 4.453855 & 4.453855 & 88.350 & 1.0 & 2943.8 & 0.01854\\
sig & 0.000650 & 0.000650 & 0.000650 & 0.039 & & 274.4 & 0.00248\\
\hline
\\
\hline
name & x & y & z & frac & Uiso \\
\hline
Mn1\\
values & 0.00000 & 0.00000 & 0.00000 & 1.000 & 0.00781\\
sig \\
O1\\
values & 0.50000 & 0.50000 & 0.50000 & 1.000 & 0.00912\\
sig \\
\end{longtable}

\clearpage

\setlength\LTleft{-0.5cm}
\begin{longtable}{c|c c c c c c c c}
\caption{Rietveld refinement parameters for the 300~K \mnexcess\ POWGEN NPD data shown in Fig. 6. The figures of merit are wR = 9.24\%, chi$^2$ = 66237.2, GOF = 4.55. The sum of all the structural phases were constrained to be 1. The following atoms were contrained to be equivalent - CuMnAs-1:Cu1 = CuMnAs-1:Mn2, CuMnAs-1:Mn1 = k=0:Mn1\_1, CuMnAs-2:Mn1 = CuMnAs-2:Cu4 = k=[001/2]:Mn1\_0, CuMnAs-2:Mn2 = CuMnAs-2:Cu5 = k=[001/2]:Mn2\_1. Crystallite size, mustrain and lattice parameters of CuMnAs-1 and k=0, CuMnAs-2 and k=[001/2] were constrained to be equal.}\\
\hline
names & Scale & Absorption\\
\hline
values & 32640.2847 & 0.0472\\
sig & 2534.3204 & 0.0027\\
\hline
\\
\hline
names & bg1 & bg2 & bg3 & bg4 & bg5 & bg6\\
\hline
values & 13.91 & -2.752 & 5.464 & 35.19 & -1.032 & -38.96\\
sig & 0.2748 & 0.6991 & 2.05 & 2.724 & 2.409 & 2.72\\
\hline
\\
& CuMnAs-1\\
\hline
names & a & b & c & Volume & Size & Mustrain & Phase fraction\\
\hline
values & 3.79264 & 3.79264 & 6.32209 & 90.938 & 0.31072 & 2909.0 & 0.70386\\
sig & & & & & 0.0426 & 75.0 & 0.06387\\
\hline
\\
\hline
names & D11 & D33\\
\hline
values & 4.924e-05 & 2.853e-06\\
sig & 3.366e-06 & 1.56e-06\\
\hline
\\
\hline
name & x & y & z & frac & Uiso \\
\hline
As1\\
values & 0.25000 & 0.25000 & 0.24027 & 1.000 & 0.01168\\
sig & & & 0.00054 & & 0.00082\\
Cu1\\
values & -0.25000 & 0.25000 & 0.50000 & 0.809 & 0.01214\\
sig & & & & 0.010 & 0.00095\\
Mn1\\
values & 0.25000 & 0.25000 & -0.16436 & 1.000 & 0.01580\\
sig & & & 0.00094 & & 0.00105\\
Mn2\\
values & -0.25000 & 0.25000 & 0.50000 & 0.191 & 0.01214\\
sig & & & & 0.010 & 0.00095\\
\hline
\\
& CuMnAs-2\\
\hline
names & a & b & c & Volume & Size & Mustrain & Phase fraction\\
\hline
values & 3.78586 & 3.78586 & 6.29098 & 90.167 & 0.49549 & 2575.7 & 0.28444\\
sig & & & & & 0.2355 & 169.7 & 0.04327\\
\hline
\\
\hline
names & D11 & D33\\
\hline
values & -9.187e-05 & -9.262e-06\\
sig & 7.711e-06 & 1.688e-06\\
\hline
\\
\hline
name & x & y & z & frac & Uiso \\
\hline
As1\\
values & 0.25000 & 0.25000 & 0.73333 & 0.808 & 0.00451\\
sig & & & 0.00170 & 0.087 & 0.00311\\
Mn1\\
values & 0.75000 & 0.25000 & 0.00000 & 0.986 & 0.00368\\
sig & & & & 0.073 & 0.00341\\
Mn2\\
values & 0.25000 & 0.25000 & 0.32711 & 0.958 & 0.01515\\
sig & & & 0.00330 & 0.121 & 0.00434\\
Cu4\\
values & 0.75000 & 0.25000 & 0.00000 & 0.014 & 0.00368\\
sig & & & & 0.073 & 0.00341 \\
Cu5\\
values & 0.25000 & 0.25000 & 0.32711 & 0.042 & 0.01515\\
sig & & & 0.00330 & 0.121 & 0.00434\\
\hline
\\
& $k = [0 0 1/2]$\\
\hline
names & a & b & c & Volume & Size & Mustrain & Phase fraction\\
\hline
values & 3.78586 & 3.78586 & 12.58195 & 180.334 & 0.49549 & 2575.7 & 0.14222\\
sig & & & & & 0.2355 & 169.7 & 0.02163\\
\hline
\\
\hline
names & D11 & D22 & D33\\
\hline
values & -9.187e-05 & -9.187e-05 & -9.262e-06\\
sig & 7.711e-06 & 7.711e-06 & 1.688e-06\\
\hline
\\
\hline
name & x & y & z & frac & Uiso \\
\hline
Mn1\_0\\
values & 0.25000 & 0.75000 & 0.00000 & 0.986 & 0.00368\\
sig & & & & 0.073 & 0.00341\\
Mn2\_1\\
values & 0.75000 & 0.75000 & 0.15647 & 0.958 & 0.01515\\
sig & & & 0.00661 & 0.121 & 0.00434 \\
\hline
\\
\hline
name & My \\
\hline
Mn1\_0\\
values & 1.382\\
sig & 0.112\\
Mn2\_1\\
values & -3.876\\
sig & 0.369\\
\hline
\\
& $k = 0$\\
\hline
names & a & b & c & Volume & Size & Mustrain & Phase fraction\\
\hline
values & 3.79264 & 3.79264 & 6.32209 & 90.938 & 0.31072 & 2909.0 & 0.70386\\
sig & & & & & 0.0426 & 75.0 & 0.06387\\
\hline
\\
\hline
names & D11 & D22 & D33\\
\hline
values & 4.924e-05 & 4.924e-05 & 2.853e-06\\
sig & 3.366e-06 & 3.366e-06 & 1.56e-06\\
\hline
\\
\hline
name & x & y & z & frac & Uiso \\
\hline
Mn1\_1\\
values & 0.25000  & 0.25000 &  0.83564  & 1.000 & 0.01580\\
sig & & & 0.00094 & & 0.00105\\
\hline
\\
\hline
name & Mx \\
\hline
Mn1\_1\\
values & 3.066\\
sig & 0.069\\
\hline
\\
& MnO\\
\hline
names & a & b & c & Volume & Size & Mustrain & Phase fraction\\
\hline
values & 4.441468 & 4.441468 & 4.441468 & 87.615 & 1.0 & 2326.9 & 0.01170\\
sig & 0.000855 & 0.000855 & 0.000855 & 0.051 & & 350.8 & 0.02062\\
\hline
\\
\hline
name & x & y & z & frac & Uiso \\
\hline
Mn1\\
values & 0.00000 & 0.00000 & 0.00000 & 1.000 & 0.00781\\
sig & & & & &\\
O1\\
values & 0.50000 & 0.50000 & 0.50000 & 1.000 & 0.00912\\
sig & & & & &\\
\end{longtable}

\clearpage

\setlength\LTleft{-1.5cm}
\begin{longtable}{c|c c c c c c c}
\caption{Rietveld refinement parameters for the room temperature \asexcess\ lab-source XRD data shown in Fig. 7(a). The figures of merit are wR = 4.72\%, chi$^2$ = 286935, GOF = 8.43.}\\
\hline
names & Scale & Shift & Transparency\\
\hline
values & 959.1755 & 34.7276 & 3.1775\\
sig & 7.1741 & 1.9226 & 0.1837\\
\hline
\\
\hline
names & bg1 & bg2 & bg3 & bg4 & bg5 & bg6\\
\hline
values & 2.74e+04 & 66.33 & 8533 & 1.425e+04 & -4.734e+04 &-1.628e+04\\
sig & 58.52 & 268.4 & 1122 & 1910 & 5432 & 3944\\
\hline
\\
\hline
names & bg7 & bg8 & bg9\\
\hline
values & 8.393e+04 & 2340 & -4.369e+04\\
sig & 9029 & 2406 & 4789\\
\hline
\\
& CuMnAs\\
\hline
names & a & b & c & Volume & Size & Mustrain & March-Dollase\\
\hline
values & 3.798740 & 3.798740 & 6.342385 & 91.523 & 0.13017 & 567.7 & 0.8171\\
sig & 0.000114 & 0.000114 & 0.000178 & 0.008 & 0.0040 & 95.8 & 0.0022\\
\hline
\\
\hline
name & x & y & z & frac & Uiso \\
\hline
As1\\
values & 0.25000 & 0.25000 & 0.22954 & 1.000 & 0.01837\\
sig & & & 0.00047 & & 0.00123\\
Cu1\\
values & -0.25000 & 0.25000 & 0.50000 & 1.000 & 0.02533\\
sig & & & & & 0.00104\\
Mn1\\
values & 0.25000 & 0.25000 & -0.16357 & 1.000 & 0.05016\\
sig & & & 0.00076 & & 0.00244\\

\end{longtable}

\clearpage

\setlength\LTleft{0.5cm}
\begin{longtable}{c|c c c c c c c}
\caption{Rietveld refinement parameters for the 520~K \asexcess\ Echidna NPD data shown in Fig. 7(b). The figures of merit are wR = 6.05\%, chi$^2$ = 17600.9, GOF = 2.41. The atomic fractions were constrained to be 1 for all atoms. The crystallite size, mustrain and the lattice parameters were constrained to be equal between the CuMnAs and k=0 phase. CuMnAs:Mn1 was constrained to be equivalent to k=0:Mn1\_1.}\\
\hline
names & Scale & DisplaceX & DisplaceY\\
\hline
values & 170953.6042 & 143.2791 & -719.8432\\
sig & 1052.1409 & 17.1238 & 19.7740\\
\hline
\\
\hline
names & bg1 & bg2 & bg3 & bg4 & bg5 & bg6\\
\hline
values & 1131 & -71.04 & 235.2 & -119.9 & 849.7 & 1710\\
sig & 4.257 & 18.84 & 82.01 & 142.5 & 408.4 & 305.3\\
\hline
\\
\hline
names & bg7 & bg8 & bg9\\
\hline
values & -3244 & -2454 & 3549\\
sig & 696.3 & 191.9 & 377.7\\
\hline
\\
& CuMnAs\\
\hline
names & a & b & c & Volume & Size & Mustrain\\
\hline
values & 3.83151 & 3.83151 & 6.33953 & 93.067 & 1.0 & 2004.2\\
sig & & & & & & 36.4\\
\hline
\\
\hline
names & D11 & D33\\
\hline
values & -2.391e-06 & -2.623e-07\\
sig & 2.663e-06 & 1.445e-06\\
\hline
\\
\hline
name & x & y & z & frac & Uiso \\
\hline
As1\\
values & 0.25000 & 0.25000 & 0.23741 & 1.000 & 0.01193\\
sig & & & 0.00046 & & 0.00133\\
Cu1\\
values & -0.25000 & 0.25000 & 0.50000 & 1.000 & 0.02023\\
sig & & & & & 0.00060\\
Mn1\\
values & 0.25000 & 0.25000 & -0.16787 & 1.000 & 0.03952\\
sig & & & 0.00096 & & 0.00238\\
\hline
\\
& $k = 0$\\
\hline
names & a & b & c & Volume & Size & Mustrain\\
\hline
values & 3.83151 & 3.83151 & 6.33953 & 93.067 & 1.0 & 2004.2\\
sig & & & & & & 36.4\\
\hline
\\
\hline
names & D11 & D22 & D33\\
\hline
values & -2.391e-06 & -2.391e-06 & -2.623e-07\\
sig & 2.663e-06 & 2.663e-06 & 1.445e-06\\
\hline
\\
\hline
name & x & y & z & frac & Uiso \\
\hline
Mn1\_1\\
values & 0.25000 & 0.25000 & 0.83213 & 1.000 & 0.03952\\
sig & & & 0.00096 & & 0.00238\\
\hline
\\
\hline
name & Mx\\
\hline
Mn1\_1\\
values & 1.478\\
sig & 0.042\\
\end{longtable}

\clearpage

\setlength\LTleft{+0.5cm}
\begin{longtable}{c|c c c c c c c}
\caption{Rietveld refinement parameters for the 4~K \asexcess\ Echidna NPD data shown in Fig. S5. The figures of merit are wR = 8.80\%, chi$^2$ = 35035.9, GOF = 3.35. The atomic fractions were constrained to be 1 for all atoms. The crystallite size, mustrain and the lattice parameters were constrained to be equal between the CuMnAs and k=0 phase. The isotropic thermal parameters were set to 0.0001.}\\
\hline
names & Scale & DisplaceX & DisplaceY\\
\hline
values & 182925.5424 & 73.5671 & -702.8654\\
sig & 932.2109 & 17.6553 & 20.2704\\
\hline
\\
\hline
names & bg1 & bg2 & bg3 & bg4 & bg5 & bg6\\
\hline
values & 892.9 & -25.66 & 197.5 & -198.1 & 886.2 & 1708\\
sig & 5.593 & 24.25 & 107.9 & 189.4 & 549.7 & 418.6\\
\hline
\\
\hline
names & bg7 & bg8 & bg9\\
\hline
values & -3577 & -2252 & 3695\\
sig & 963.5 & 272 & 539.1\\
\hline
\\
& CuMnAs\\
\hline
names & a & b & c & Volume & Size & Mustrain\\
\hline
values & 3.77883 & 3.77883 & 6.33191 & 90.417 & 1.0 & 2375.8\\
sig & & & & & & 36.5\\
\hline
\\
\hline
names & D11 & D33\\
\hline
values & -1.655e-06 & 1.152e-06\\
sig & 2.762e-06 & 1.476e-06\\
\hline
\\
\hline
name & x & y & z & frac & Uiso\\
\hline
As1\\
values & 0.25000 & 0.25000 & 0.23435 & 1.000 & 0.00010\\
sig & & & 0.00038\\
Cu1\\
values & -0.25000 & 0.25000 & 0.50000 & 1.000 & 0.00010\\
sig & & & & &\\
Mn1\\
values & 0.25000 & 0.25000 & -0.16259 & 1.000 & 0.00010\\
sig & & & 0.00079\\
\hline
\\
& $k = 0$\\
\hline
names & a & b & c & Volume & Size & Mustrain\\
\hline
values & 3.77883 & 3.77883 & 6.33191 & 90.417 & 1.0 & 2375.8\\
sig & & & & & & 36.5\\
\hline
\\
\hline
names & D11 & D22 & D33\\
\hline
values & -1.655e-06 & -1.655e-06 & 1.152e-06\\
sig & 2.762e-06 & 2.762e-06 & 1.476e-06\\
\hline
\\
\hline
name & x & y & z & frac & Uiso \\
\hline
Mn1\_1\\
values & 0.25000 & 0.75000 & 0.83741 & 1.000 & 0.00010\\
sig & & & 0.00079\\
\hline
\\
\hline
name & My\\
\hline
Mn1\_1\\
values & 3.733\\
sig & 0.029\\
\end{longtable}